\documentstyle{article}

\newcommand{\ba}{\begin{eqnarray}}
\newcommand{\ea}{\end{eqnarray}}
\input{psfig}
\begin{document}
\pagestyle{plain}

\title{Algebraic Models of Hadron Structure \\
II. Strange Baryons}
\author{R. Bijker\\
Instituto de Ciencias Nucleares, U.N.A.M.,\\
A.P. 70-543, 04510 M\'exico D.F., M\'exico 
\and
F. Iachello\\
Center for Theoretical Physics, Sloane Laboratory,\\
Yale University, New Haven, CT 06520-8120, U.S.A.
\and
A. Leviatan\\ 
Racah Institute of Physics, The Hebrew University,\\
Jerusalem 91904, Israel}
\date{April 3, 2000}
\maketitle

\begin{abstract}
The algebraic treatment of baryons is extended to strange 
resonances. Within this framework we study a collective 
string-like model in which the radial excitations are 
interpreted as rotations and vibrations of the strings. 
We derive a mass formula and closed expressions for 
strong and electromagnetic decay widths and use these 
to analyze the available experimental data.
\end{abstract}

\begin{center}
\end{center}

\vspace{1cm}

\begin{center}
Annals of Physics (N.Y.), in press
\end{center}

\clearpage

\section{Introduction}
\setcounter{equation}{0}

In the last few years there has been renewed interest in hadron 
spectroscopy. Especially the development of dedicated experimental 
facilities to probe the structure of hadrons in the nonperturbative 
region of QCD with far greater precision than before has generated 
a considerable amount of experimental and theoretical activity 
\cite{baryons98}. 
This has stimulated us to reexamine hadron spectroscopy in a 
novel approach in which both internal (spin-flavor-color) and 
space degrees of freedom of hadrons are treated algebraically. 
The new ingredient is the introduction of a space symmetry or 
spectrum generating algebra for the radial excitations which 
for mesons was taken as $U(4)$ \cite{meson} and for baryons as 
$U(7)$ \cite{BIL}. The algebraic approach unifies the harmonic 
oscillator quark model, $U(4) \supset U(3)$ for mesons and 
$U(7) \supset U(6)$ for baryons, with collective string-like 
models of hadrons. 

In the first paper of this series \cite{BIL} we have introduced 
$U(7)$ to study the properties of nonstrange baryons, such as 
the mass spectrum, electromagnetic couplings \cite{emff} and 
strong decays \cite{strong}. In this article we extend these 
studies to hyperons and present a systematic study of both nonstrange 
and strange baryons in the framework of a collective string-like 
$qqq$ model in which the orbital excitations are treated as 
rotations and vibrations of the strings. The algebraic structure 
of the model enables us to obtain transparent results (mass formula, 
selection rules and decay widths for strong and electromagnetic 
couplings) that can be used to analyze and interpret the 
experimental data, and look for evidence of the existence of 
unconventional ({\it i.e.} non $qqq$) configurations of quarks 
and gluons, such as hybrid quark-gluon states $qqq$-$g$ or 
multiquark meson-baryon bound states $qqq$-$q\overline{q}$. 

In particular, we discuss the mass spectrum (Sects. 3-4), the 
strong (Sects. 5-6) and electromagnetic (Sects. 7-8) decay widths. 
We do this in a framework in which spin-flavor symmetry is broken 
in a diagonal way in the masses 
({\it i.e.} through a dynamic symmetry). This asssumption 
appears to be sufficient to describe most observables. 
The breaking of spin-flavor symmetry in hyperon decays can be 
investigated using a procedure similar to that in Ref.~\cite{emff} 
for nonstrange baryons. The results of such a study will be 
published separately. 


\section{Algebraic models of baryons}
\setcounter{equation}{0}

We consider baryons to be built of three constituent parts 
which are characterized by both internal and spatial degrees 
of freedom. 

\subsection{Degrees of freedom} 

\begin{figure}
\centerline{\hbox{
\psfig{figure=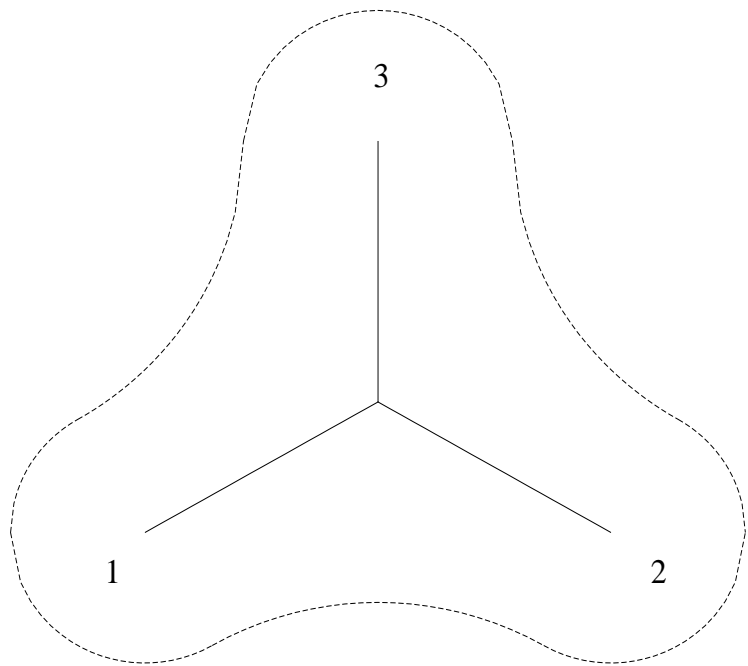,height=0.60\textwidth,width=0.8\textwidth} }}
\caption[]{Collective model of baryons.}
\label{geometry}
\end{figure}

The internal degrees of freedom of these three parts are taken to be:
flavor-triplet $u,d,s$ (for the light quark flavors), 
spin-doublet $S=1/2$, and color-triplet. The internal algebraic 
structure of the constituent parts consists of the usual spin-flavor 
and color algebras 
\ba 
{\cal G}_{\rm i} &=& SU_{\rm sf}(6) \otimes SU_{\rm c}(3) ~.
\label{sfc}
\ea 
In \cite{BIL} we discussed various algebraic models of baryons. 
These models share a common spin-flavor structure (see Eq.~(\ref{sfc}), 
but differ in their treatment of radial excitations. Here we 
consider a collective string-like model with the configuration
depicted in Fig.~\ref{geometry}. The relevant degrees of freedom 
for the relative motion of the three constituent parts of this 
configuration are provided by the relative Jacobi coordinates
which we choose as \cite{fbs}
\ba
\vec{\rho} &=& \frac{1}{\sqrt{2}} \, (\vec{r}_1 - \vec{r}_2) ~,
\nonumber\\
\vec{\lambda} &=& 
\frac{1}{\sqrt{m_1^2+m_2^2+(m_1+m_2)^2}} \,
[ m_1 \vec{r}_1 + m_2 \vec{r}_2 - (m_1+m_2) \vec{r}_3 ] ~.
\label{jacobi_gen}
\ea
Here $m_i$ and $\vec{r}_i$ ($i=1,2,3$) denote the mass and 
coordinate of the $i$-th constituent. When two of the constituents 
have equal mass ($m_1=m_2$), the above choice reduces to
\ba
\vec{\rho} &=& \frac{1}{\sqrt{2}} (\vec{r}_1 - \vec{r}_2) ~,
\nonumber\\
\vec{\lambda} &=& \frac{1}{\sqrt{6}} (\vec{r}_1 + \vec{r}_2 -2\vec{r}_3) ~.
\label{jacobi}
\ea
Since the quark masses satisfy to a good approximation 
$m_u=m_d \neq m_s$, the Jacobi coordinates of Eq.~(\ref{jacobi}) 
are relevant for all baryons be it with strangeness $S=0$, $-1$, 
$-2$ or $-3$. Instead of a formulation in terms of coordinates 
and momenta, we use the method of bosonic quantization 
in which we introduce a dipole boson with $L^P=1^-$ for each independent
relative coordinate, and an auxiliary scalar boson with $L^P=0^+$ 
\cite{BIL} 
\ba 
b^{\dagger}_{\rho,m} ~, \; b^{\dagger}_{\lambda,m} ~, \; 
s^{\dagger} \hspace{1cm} (m=-1,0,1) ~. \label{bb}
\ea
The scalar boson does not represent an independent degree of freedom, 
but is added under the restriction that the total number of bosons 
$N$ is conserved. This procedure leads to a compact spectrum generating 
algebra for the radial (or orbital) excitations 
\ba 
{\cal G}_r &=& U(7) ~.  \label{u7}
\ea 
The $U(7)$ algebra enlarges the $U(6)$ algebra of the harmonic 
oscillator quark model \cite{hoqm}, but still describes the dynamics 
of two vectors. For a system of interacting bosons the model space is 
spanned by the symmetric irreducible representation $[N]$ of $U(7)$. 
This representation contains all oscillator shells with
$n=n_{\rho}+n_{\lambda}=0,1,2,\ldots, N$. The value of $N$ determines
the size of the model space and, in view of confinement, is expected 
to be large. 

\subsection{Basis states}

The full algebraic structure is obtained by combining the spatial 
part ${\cal G}_{\rm r}$ of Eq.~(\ref{u7}) with the internal 
spin-flavor-color part ${\cal G}_{\rm i}$ of Eq.~(\ref{sfc})
\ba
{\cal G} &=& {\cal G}_{\rm r} \otimes SU_{\rm sf}(6) \otimes SU_{\rm c}(3) ~.
\ea
The spatial part of the baryon wave function has to be combined 
with the spin-flavor and color part, in such a way that the total 
wave function is antisymmetric. Since the color part of the wave 
function is antisymmetric (color singlet), the remaining part 
(space-spin-flavor) has to be symmetric. 
A convenient set of basis states is provided by the case of three 
identical constituents, for which the spatial and spin-flavor parts 
of the baryon wave function are in addition labeled by their 
transformation properties under the permutation group $S_3$: 
$t=S$ for the symmetric, $t=A$ for the antisymmetric and $t=M$ for 
the two-dimensional $(M_{\rho},M_{\lambda})$ mixed symmetry 
representation. 

A set of basis states for the spin-flavor part is provided 
by the decomposition of $SU_{\rm sf}(6)$ into its flavor and spin parts 
\ba
\left| \begin{array}{ccccccccccc}
SU_{\rm sf}(6) &\supset& SU_{\rm f}(3) &\otimes& SU_{\rm s}(2) 
&\supset& SU_{\rm I}(2) &\otimes& U_{\rm Y}(1) &\otimes&
SU_{\rm s}(2) \\
\downarrow && \downarrow && \downarrow && \downarrow && \downarrow && \\
\, [f_1 f_2 f_3] && [g_1 g_2] && S && I && Y && 
\end{array} \right> ~. \label{sfbasis}
\ea
Here $[f_1 f_2 f_3]$ and $[g_1 g_2]$ represent the Young tableaux, 
$S$ denotes the spin, $I$ the isospin and $Y$ the hypercharge. 
The representations of the spin-flavor groups are often labeled  
by their dimensions (rather than by their Young tableaux) 
\ba
\mbox{dim}[f_1 f_2 f_3] &=& 
\frac{(f_1-f_2+1)(f_1-f_3+2)(f_2-f_3+1)(f_1+5)!(f_2+4)!(f_3+3)!} 
{3!4!5!(f_1+2)!(f_2+1)!f_3!} ~, 
\nonumber\\
\mbox{dim}[g_1 g_2] &=& \frac{1}{2}(g_1-g_2+1)(g_1+2)(g_2+1) ~, 
\nonumber\\
\mbox{dim}[S] &=& 2S+1 ~.
\ea
For three constituent parts the allowed values of $[f_1 f_2 f_3]$ 
are $[300]$ ($t=S$), $[210]$ ($t=M$) and $[111]$ ($t=A$) with dimensions 
56, 70 and 20, respectively. 
The flavor part is characterized by $[g_1 g_2]=[30]$, 
$[21]$ or $[00]$ with dimensions 10 (decuplet), 8 (octet) 
or 1 (singlet), respectively. In the 
notation of \cite{deSwart} the flavor wave functions are labeled 
by $(p,q)=(g_1-g_2,g_2)$. Finally, the total spin of three spin-1/2 
objects is $S=3/2$ or $S=1/2$. 
The decomposition of representations of 
$SU_{\rm sf}(6)$ into those of $SU_{\rm f}(3) \otimes SU_{\rm s}(2)$
is the standard one  
\ba 
\, S \leftrightarrow [56] &\supset& ^{2}8 \, \oplus \, ^{4}10 ~,
\nonumber\\
\, M \leftrightarrow [70] &\supset& 
^{2}8 \, \oplus \, ^{4}8 \, \oplus \, ^{2}10 \, \oplus \, ^{2}1 ~,
\nonumber\\
\, A \leftrightarrow [20] &\supset& ^{2}8 \, \oplus \, ^{4}1 ~, 
\ea
where we have denoted the irreducible representations by their 
dimensions. Each flavor multiplet consists of families of baryons which 
are characterized by their isospin $I$ and hypercharge $Y$ 
(see Table~\ref{iy}). 
The electric charge is given by the Gell-Mann and Nishijima relation 
\ba
Q &=& I_3 + \frac{Y}{2} ~.
\ea
In Appendix~A we present the explicit form for the spin and 
flavor wave functions in the convention that we have used in 
this paper. 

Since the space-spin-flavor wave function 
is symmetric ($t=S$), the symmetry of the spatial wave function 
under $S_3$ has to be the same as that of the spin-flavor part. 
Hence it is convenient to label the spatial wave functions by 
the basis states of a dynamical symmetry of $U(7)$ that preserves 
the $S_3$ permutation symmetry. We choose the chain that correponds 
to the problem of three particles in a common harmonic oscillator 
potential \cite{KM}
\ba
\left| \begin{array}{ccccccccccc}
U(7) &\supset& U(6) &\supset& {\cal SU}(3) &\otimes& SU(2) 
&\supset& {\cal SO}(3) &\otimes& SO(2) \\
N &,& n &,& (n_1,n_2) &,& F &,& L &,& m_F \end{array} \right> ~. 
\label{ch1}
\ea
In this decomposition, the behavior in three-dimensional coordinate 
space ${\cal SU}(3) \supset {\cal SO}(3)$ is separated from that in 
index space $SU(2) \supset SO(2)$. 
The allowed values of the quantum numbers can be obtained from the 
branching rules. For the decomposition of $U(6)$ we use the 
complementarity relationship between 
the groups ${\cal SU}(3)$ and $SU(2)$ within the symmetric 
irreducible representation $U(6)$. As a consequence, the labels 
of ${\cal SU}(3)$ are determined by those of $SU(2)$. 
The branching rules are 
\ba
n &=& 0,1,\ldots,N ~, 
\nonumber\\
F &=& n,n-2,\ldots,1 \mbox{ or } 0 ~,
\nonumber\\ 
(n_1,n_2) &=& (\frac{n+F}{2},\frac{n-F}{2}) ~, 
\nonumber\\
m_F &=& -F,-F+2,\ldots,F ~. \label{br1}
\ea
The reduction from the coupled harmonic oscillator group 
to the rotation group ${\cal SU}(3) \supset {\cal SO}(3)$ 
is given by \cite{Elliott}
\ba
K &=& \mbox{min}\{\lambda,\mu\},\mbox{min}\{\lambda,\mu\}-2,\ldots,
1 \mbox{ or } 0 ~,
\nonumber\\
K &=& 0: \hspace{1cm} L \;=\; \mbox{max}\{\lambda,\mu\},
\mbox{max}\{\lambda,\mu\}-2,\ldots,1 \mbox{ or } 0 ~, 
\nonumber\\
K &>& 0: \hspace{1cm} 
L \;=\; K,K+1,\ldots,K+\mbox{max}\{\lambda,\mu\} ~. 
\label{JPE} 
\ea
Here $(\lambda,\mu)=(n_1-n_2,n_2)=(F,(n-F)/2)$. The label $K$ is 
an extra label that has to be introduced to label the states 
uniquely \cite{Elliott}. 
The $SO(2)$ group in Eq.~(\ref{ch1}) is related to the permutation 
symmetry \cite{BIL,KM,hoqm}. The states with good $S_3$ symmetry 
are given by the linear combinations 
\ba
| \psi_1 \rangle &=& \frac{-i}{\sqrt{2(1+\delta_{m_F,0})}} \, 
\left[ | \phi_{m_F} \rangle - | \phi_{-m_F} \rangle \right] ~,
\nonumber\\
| \psi_2 \rangle &=& \frac{(-1)^{\nu}}{\sqrt{2(1+\delta_{m_F,0})}} \, 
\left[ | \phi_{m_F} \rangle + | \phi_{-m_F} \rangle \right] ~. 
\label{wfp12}
\ea
Here we have introduced the label $\nu$ by $m_F=\nu$ (mod 3). 
The wave functions $| \psi_1 \rangle$ 
($| \psi_2 \rangle$) transform for $\nu=0$ as $t=A$ ($S$), and 
for $\nu=1,2$ as $t=M_{\rho}$ ($M_{\lambda}$).
Summarizing, the basis states are characterized uniquely by 
\ba
| N,n,F,m_F,K,L^P_t \rangle ~, \label{basis1}
\ea
where $P$ is the parity of the basis states $P=(-)^n$. 
Finally, the quark orbital angular momentum $L$ is coupled 
with the spin $S$ to the total angular momentum $J$ of the 
baryon. 
In Appendix~B we present the space-spin-flavor baryon wave 
functions with $S_3$ symmetry. 


\section{Mass operator}
\setcounter{equation}{0}

The mass operator depends both on the spatial and the internal 
degrees of freedom. For the spatial part we adopt a collective 
model of the nucleon in which the baryons are interpreted as 
rotational and vibrational excitations of the string configuration 
of Fig.~\ref{geometry}. For two identical constituent parts 
(as is the case for strange baryons) the vibrations are 
described by \cite{fbs}
\ba
\hat{M}^2_{\rm vib} &=& A \, P_1^{\dagger} P_1 + 
B \, P_2^{\dagger} P_2 + C \, P_3^{\dagger} P_3 
+ D \, ( P_1^{\dagger} P_2 + P_2^{\dagger} P_1 ) ~, 
\label{mvib1}
\ea
with
\ba
P_1^{\dagger} &=& R^2 \, s^{\dagger} s^{\dagger} 
- b^{\dagger}_{\rho} \cdot b^{\dagger}_{\rho}
- b^{\dagger}_{\lambda} \cdot b^{\dagger}_{\lambda} ~,
\nonumber\\
P_2^{\dagger} &=&   (\cos \beta)^2 \, 
b^{\dagger}_{\rho} \cdot b^{\dagger}_{\rho} - (\sin \beta)^2 \, 
b^{\dagger}_{\lambda} \cdot b^{\dagger}_{\lambda} ~, 
\nonumber\\
P_3^{\dagger} &=& b^{\dagger}_{\rho} \cdot b^{\dagger}_{\lambda} ~. 
\label{mvib2}
\ea
Here $R$ is related to the hyperspherical radius 
$\sqrt{\rho^2+\lambda^2}$, and $\beta$ corresponds to the 
hyperspherical angle $\tan\beta = \rho/\lambda$ with 
$\rho=|\vec{\rho}|$ and $\lambda=|\vec{\lambda}|$. 
The mass operator in this case is $S_2$ invariant. 
In the limit of a large model space ($N \rightarrow \infty$) the
mass operator of Eqs.~(\ref{mvib1})-(\ref{mvib2}) reduces to leading
order in $N$ to a harmonic form, and its eigenvalues are given by 
\cite{fbs} 
\ba
M^2_{\rm vib} &=& \kappa_1 \, n_u + \kappa_2 \, n_v + \kappa_3 \, n_w ~,
\ea
Here $\kappa_1$, $\kappa_2$ are the eigenvalues of the 
$2\times2$ symmetric matrix 
\ba
\left( \begin{array}{cc} 4ANR^2 & 
2DN\sin(2\beta)R^2/\sqrt{1+R^2} \\ 
2DN\sin(2\beta)R^2/\sqrt{1+R^2} & 
BN\sin^{2}(2\beta)R^2/(1+R^2) \end{array} \right) ~. 
\ea
and $\kappa_3=CNR^2/(1+R^2)$. The vibrational quantum numbers 
$n_u$, $n_v$ and $n_w$ denote the number of quanta in the symmetric 
stretching (or breathing mode), antisymmetric stretching and bending 
vibrations of the strings, respectively (see Fig.~3 of \cite{BIL}). 
For three 
identical constituents we obtain the $S_3$-invariant mass operator 
of \cite{BIL} from Eqs.~(\ref{mvib1})-(\ref{mvib2}) by taking 
$D=0$, $B=C$ and $\beta=\pi/4$, which leads to 
$\kappa_1 = 4ANR^2$ and $\kappa_2 = \kappa_3 = BNR^2/(1+R^2)$. 
In the analysis of the mass spectrum of strange baryons, to be 
presented below, the $S_3$ symmetry of the mass operator is only 
broken dynamically in the spin-flavor part. Therefore, the baryon 
wave functions still have good $S_3$ symmetry, and the vibrational 
part of the mass operator only depends on $\kappa_1$ and 
$\kappa_2$ for all baryons 
\ba
M^2_{\rm vib} &=& \kappa_1 \, v_1 + \kappa_2 \, v_2 ~.
\ea
Here $v_1=n_u$ and $v_2=n_v+n_w$ are the vibrational quantum numbers 
corresponding to the symmetric stretching vibration along the 
direction of the strings (breathing mode), and two degenerate 
bending vibrations of the strings. The spectrum consists of a 
series of vibrational excitations characterized by the labels 
$(v_1,v_2)$, and a tower of rotational excitations built on top 
of each vibration. The occurrence of linear 
Regge trajectories suggests to add a term linear in $L$ to the 
mass operator
\ba
M^2_{\rm space} &=& \kappa_1 \, v_1 + \kappa_2 \, v_2 + \alpha \, L ~.
\label{mrad}
\ea
In the application to nonstrange baryons \cite{BIL} the Roper N$(1440)$,
the $\Delta(1600)$ and the $\Delta(1900)$ resonances were assigned 
to the symmetric stretching vibration $(v_1,v_2)=(1,0)$,
and the N$(1710)$ resonance to the $(v_1,v_2)=(0,1)$ vibration.
The remaining resonances were interpreted as rotational excitations. 

For the spin-flavor part of the mass operator we use the 
G\"ursey-Radicati \cite{GR} form
\ba
\hat M^2_{\rm sf}
&=& a \, \Bigl[ \hat C_2(SU_{\rm sf}(6)) - 45 \Bigr]
  + b \, \Bigl[ \hat C_2(SU_{\rm f}(3)) -  9 \Bigr]
  + c \, \Bigl[ \hat C_2(SU_{\rm s}(2)) - \frac{3}{4} \Bigr] 
\nonumber\\
&& + d \, \Bigl[ \hat C_1(U_{\rm Y}(1)) - 1 \Bigr]
   + e \, \Bigl[ \hat C_2(U_{\rm Y}(1)) - 1 \Bigr] 
   + f \, \Bigl[ \hat C_2(SU_{\rm I}(2)) - \frac{3}{4} \Bigr] ~. 
\label{msf}
\ea
The eigenvalues of the Casimir operators in the basis states of 
Eq.~(\ref{sfbasis}) are
\ba
\langle \hat C_2(SU_{\rm sf}(6)) \rangle &=& 
2\left[ f_1(f_1+5) + f_2(f_2+3) + f_3(f_3+1) 
- \frac{1}{6} (f_1+f_2+f_3)^2 \right] ~, 
\nonumber\\
\langle \hat C_2(SU_{\rm f}(3)) \rangle &=& 
\frac{3}{2} \left[ g_1(g_1+2) + g_2^2 
- \frac{1}{3} (g_1+g_2)^2 \right] ~, 
\nonumber\\
\langle \hat C_2(SU_{\rm s}(2)) \rangle &=& S(S+1) ~,
\nonumber\\
\langle \hat C_1(U_{\rm Y}(1)) \rangle &=& Y ~,
\nonumber\\
\langle \hat C_2(U_{\rm Y}(1)) \rangle &=& Y^2 ~,
\nonumber\\
\langle \hat C_2(SU_{\rm I}(2)) \rangle &=& I(I+1) ~. 
\label{sfcasimir}
\ea
We have defined the operators such that each of the terms vanishes
for the ground state of the nucleon. 
The spin term represents spin-spin interactions, the flavor term
denotes the flavor dependence of the interactions, and the
$SU_{\rm sf}(6)$ term, which according to Eq.~(\ref{sfcasimir})
depends on the permutation symmetry of the wave functions, represents
`signature dependent' interactions. These signature dependent (or 
exchange) interactions were extensively investigated years ago within 
the framework of Regge theory \cite{exchange}. 
The last two terms represent the isospin and hypercharge dependence 
of the masses. We do not consider here interaction terms that mix 
the space and internal degrees of freedom.


\section{Comparison with experimental mass spectrum}
\setcounter{equation}{0}

In this section we analyze simultaneously the experimental mass spectrum 
of strange and nonstrange baryons in terms of the mass formula 
\ba
M^2 &=& M^2_0 + \kappa_1 \, v_1 + \kappa_2 \, v_2 + \alpha \, L 
\nonumber\\ 
&& + a \, \Bigl[ 2f_1(f_1+5) + 2f_2(f_2+3) + 2f_3(f_3+1) 
- \frac{1}{3} (f_1+f_2+f_3)^2 - 45 \Bigr] 
\nonumber\\ 
&& + b \, \Bigl[ \frac{3}{2} \left( g_1(g_1+2) + g_2^2 
- \frac{1}{3} (g_1+g_2)^2 \right) - 9 \Bigr] 
   + c \, \Bigl[ S(S+1) - \frac{3}{4} \Bigr] 
\nonumber\\ 
&& + d \, \Bigl[ Y - 1 \Bigr]
   + e \, \Bigl[ Y^2 - 1 \Bigr] 
   + f \, \Bigl[ I(I+1) - \frac{3}{4} \Bigr] ~.
\label{massformula}
\ea
The coefficient $M^2_0$ is determined by the nucleon mass 
$M_{0}^{2}=0.882$ GeV$^2$. The remaining nine coefficients are 
obtained in a simultaneous fit to the three and four star 
resonances of Tables~\ref{nucdel} and \ref{strange} which have 
been assigned as octet and decuplet states. We find a good overall 
fit for 48 resonances with an r.m.s. deviation of $\delta=33$ MeV. 
The values of the parameters are given in Table~\ref{par}. 
In the last column we show for comparison the parameters 
that were obtained in a fit to 25 nucleon and delta 
resonances with a r.m.s. deviation to $\delta=39$ MeV \cite{BIL}. 
In comparison with Table~II of \cite{BIL} the $\Delta(1900)S_{31}$ 
was left out, since it has been downgraded from a three to a two 
star resonance \cite{PDG}. Since for nonstrange resonances $Y=1$, 
the $d$ and $e$ terms in Eq.~(\ref{massformula}) do not 
contribute. The flavor and isospin dependent terms that 
determine the mass splitting between the nucleon and $\Delta$ 
resonances can be combined into a single $b$ term with strength 
$b+\frac{1}{3}f = 0.031$ GeV$^2$, very close to the fitted value 
of $0.030$ GeV$^2$ for nonstrange baryons. Thus, the parameter 
values determined in the present simultaneous study of both 
strange and nonstrange resonances are almost the same as those 
found for nonstrange resonances.  

Tables~\ref{nucdel} and \ref{strange} and 
Figs.~\ref{nstar}--\ref{lstar} show that the mass formula of 
Eq.~(\ref{massformula}) provides a good overall description of both 
positive and negative baryon resonances belonging to the $N$,  
$\Delta$, $\Sigma$, $\Lambda$, $\Xi$ and $\Omega$ families. 
There is no need for an additional energy shift for the positive 
parity states and another one for the negative parity states, 
as in the relativized quark model \cite{rqm}. 

\subsection{Octet and decuplet resonances}

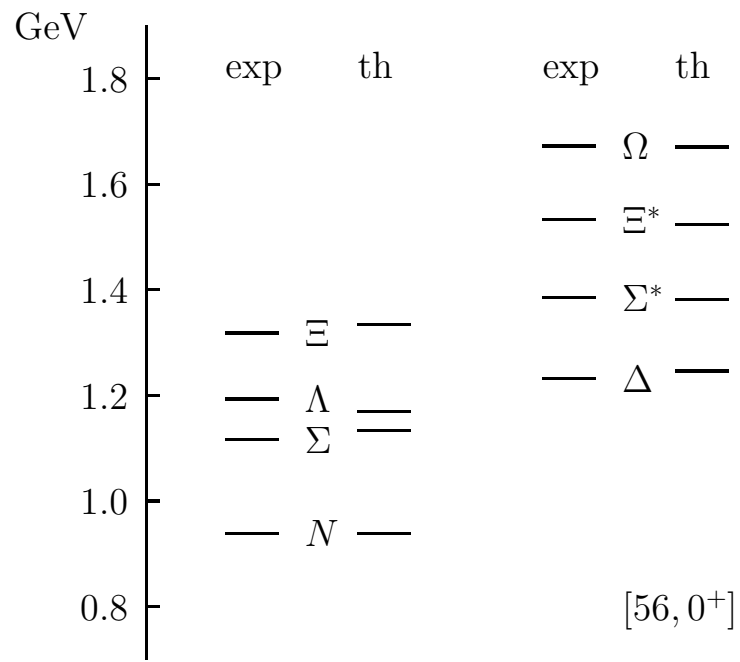
\begin{figure}
\centering
\Large
\setlength{\unitlength}{1pt}
\begin{picture}(280,280)(0,0)
\thicklines
\put ( 50, 20) {\line(0,1){240}}
\put ( 50, 40) {\line(1,0){5}}
\put ( 50, 80) {\line(1,0){5}}
\put ( 50,120) {\line(1,0){5}}
\put ( 50,160) {\line(1,0){5}}
\put ( 50,200) {\line(1,0){5}}
\put ( 50,240) {\line(1,0){5}}
\put ( 25, 35) {0.8}
\put ( 25, 75) {1.0}
\put ( 25,115) {1.2}
\put ( 25,155) {1.4}
\put ( 25,195) {1.6}
\put ( 25,235) {1.8}
\put (  0,255) {GeV}
\put ( 80, 67.8) {\line(1,0){20}}
\put ( 80,103.2) {\line(1,0){20}}
\put ( 80,118.6) {\line(1,0){20}}
\put ( 80,143.6) {\line(1,0){20}}
\put (110, 62.8) {$N$}
\put (110, 98.2) {$\Sigma$}
\put (110,113.6) {$\Lambda$}
\put (110,138.6) {$\Xi$}
\put (130, 67.8) {\line(1,0){20}}
\put (130,106.6) {\line(1,0){20}}
\put (130,114.0) {\line(1,0){20}}
\put (130,146.8) {\line(1,0){20}}
\put (200,126.4) {\line(1,0){20}}
\put (200,157.0) {\line(1,0){20}}
\put (200,186.6) {\line(1,0){20}}
\put (200,214.4) {\line(1,0){20}}
\put (230,121.4) {$\Delta$}
\put (230,152.0) {$\Sigma^{*}$}
\put (230,181.6) {$\Xi^{*}$}
\put (230,209.4) {$\Omega$}
\put (250,129.2) {\line(1,0){20}}
\put (250,156.4) {\line(1,0){20}}
\put (250,184.8) {\line(1,0){20}}
\put (250,214.0) {\line(1,0){20}}
\put ( 80,240) {exp}
\put (130,240) {th}
\put (200,240) {exp}
\put (250,240) {th}
\put (230, 35) {$[56,0^+]$}
\end{picture}
\normalsize
\vspace{1cm}
\caption[]{Ground state baryon octet with $J^P=1/2^+$ 
and decuplet with $J^P=3/2^+$}
\label{gsbar}
\end{figure}

\begin{figure}
\centerline{\hbox{\psfig{figure=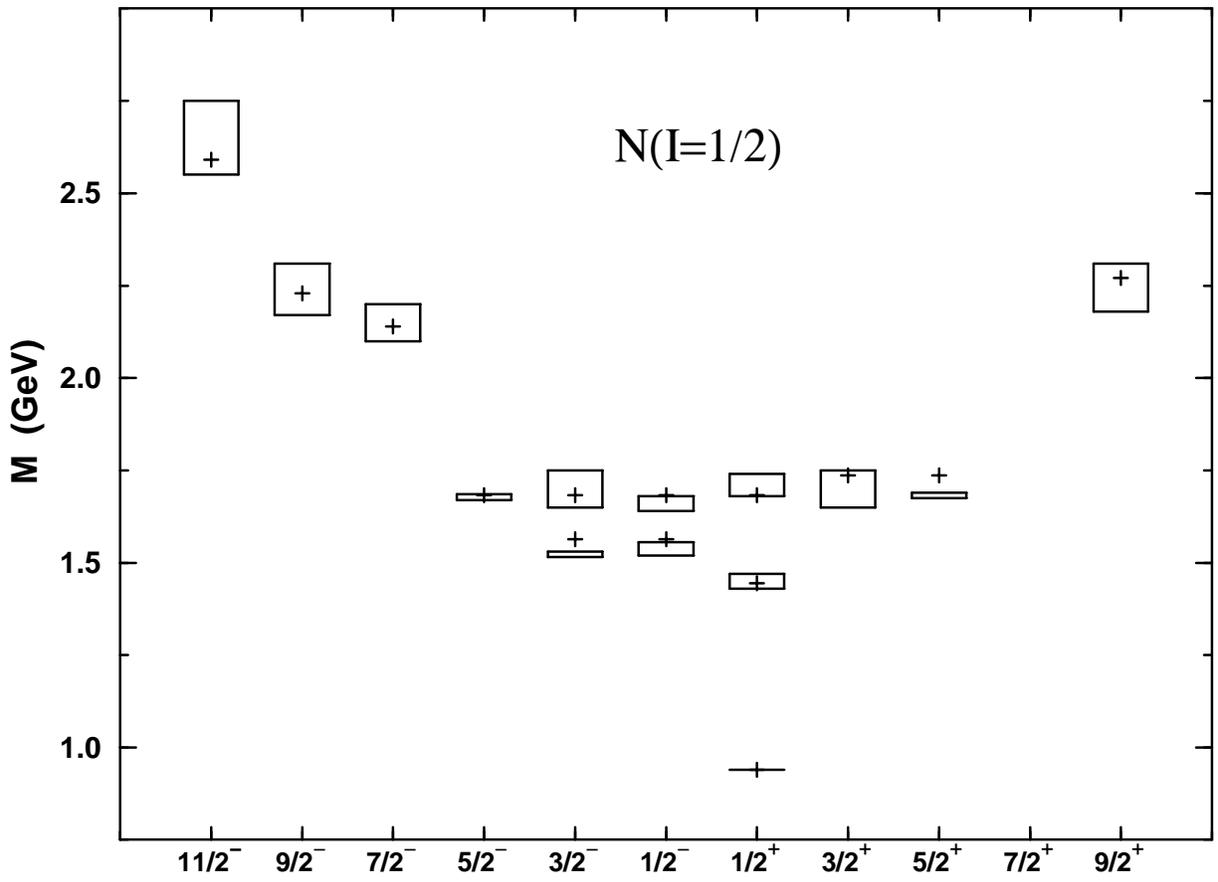,height=0.60\textwidth,width=0.60\textwidth,angle=270} }}
\caption[]{Comparison between the experimental mass spectrum 
of three and four star nucleon resonances (boxes) and the 
calculated masses ($+$).} 
\label{nstar}
\end{figure}

\begin{figure}
\centerline{\hbox{\psfig{figure=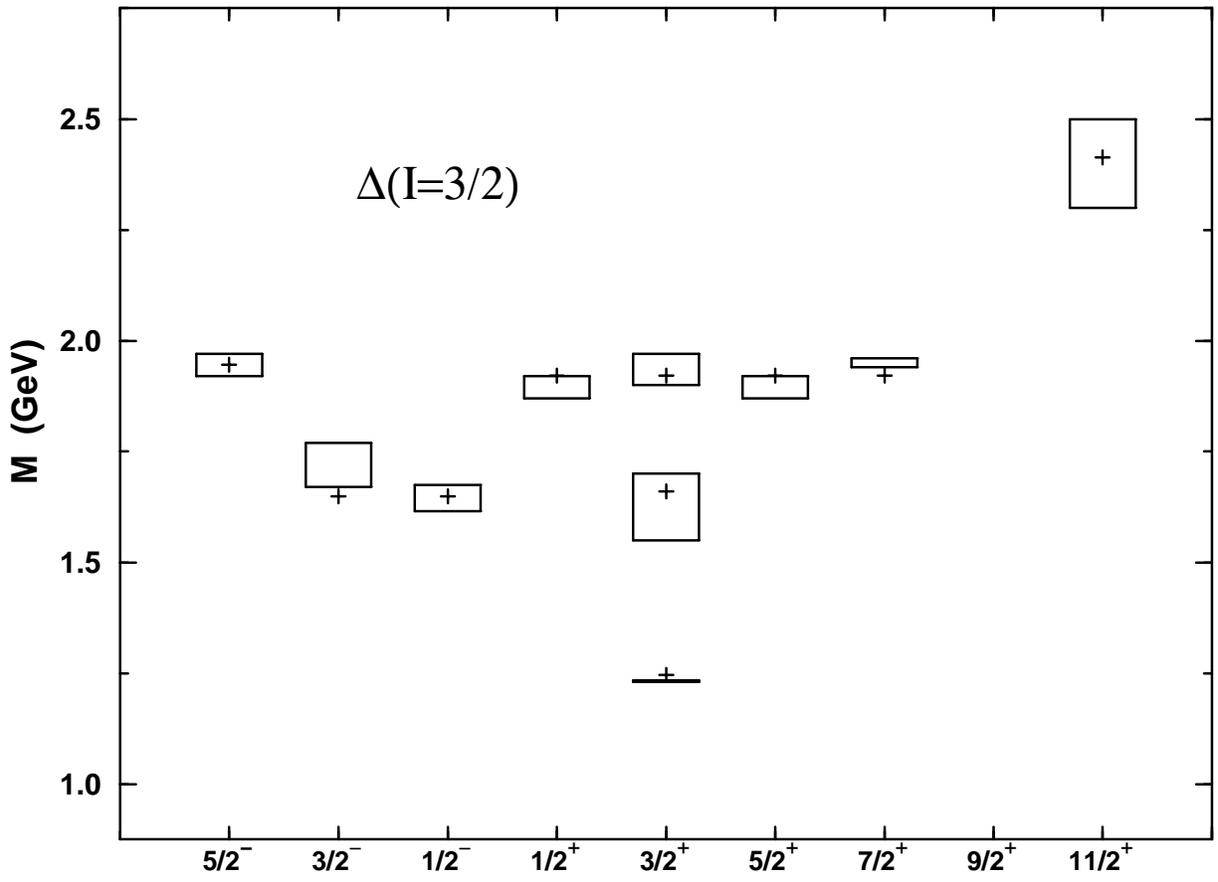,height=0.60\textwidth,width=0.60\textwidth,angle=270} }}
\caption[]{As Fig.~\ref{nstar}, but for $\Delta$ resonances.}
\label{dstar}
\end{figure}

\begin{figure}
\centerline{\hbox{\psfig{figure=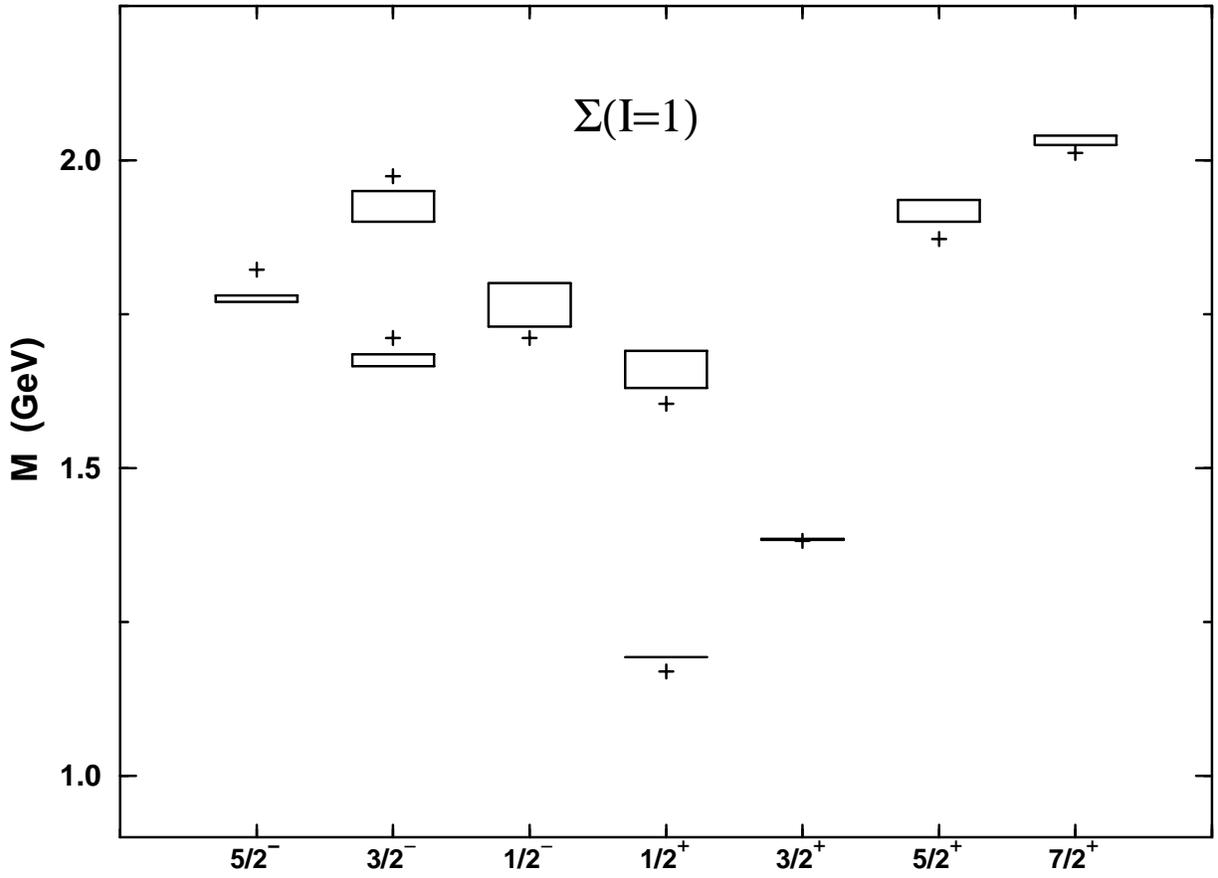,height=0.60\textwidth,width=0.60\textwidth,angle=270} }}
\caption[]{As Fig.~\ref{nstar}, but for $\Sigma$ resonances.}
\label{sstar}
\end{figure}

\begin{figure}
\centerline{\hbox{\psfig{figure=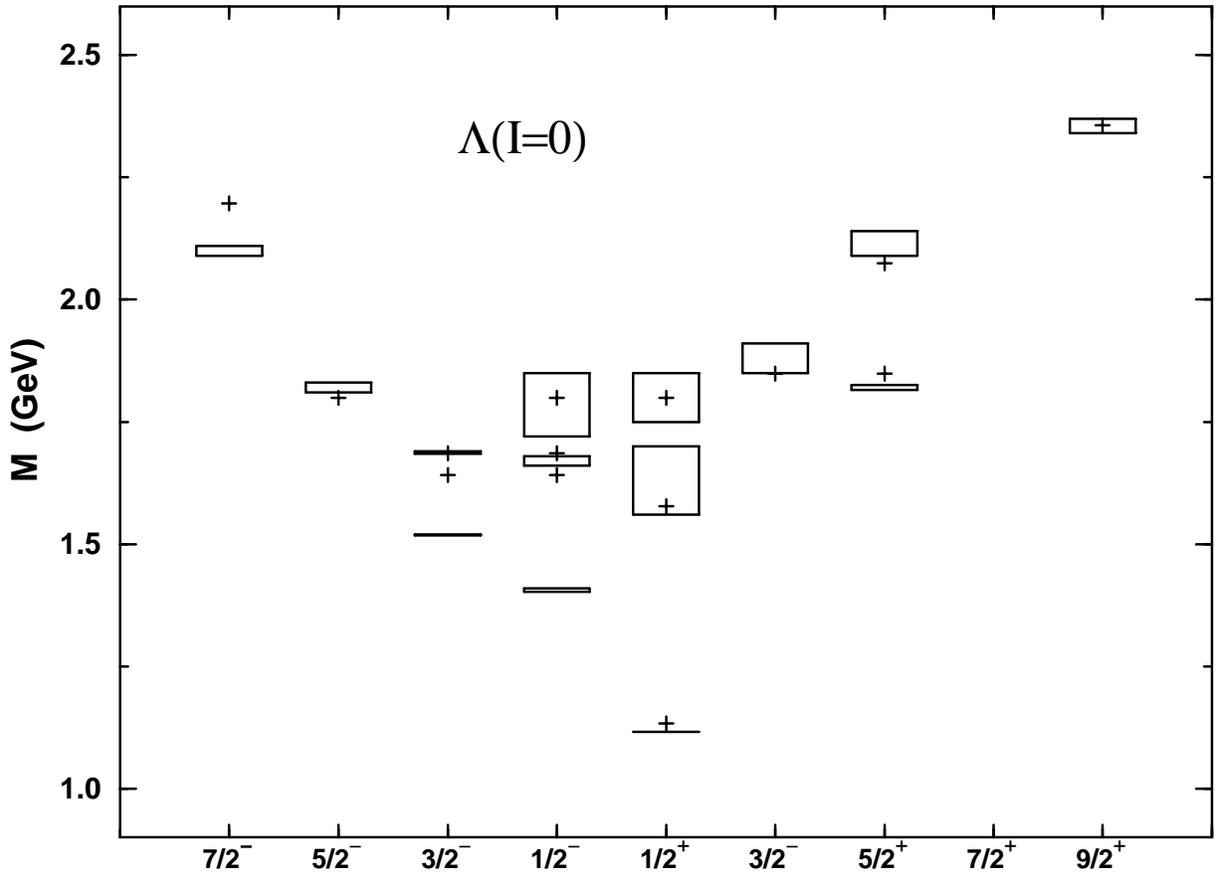,height=0.60\textwidth,width=0.60\textwidth,angle=270} }}
\caption[]{As Fig.~\ref{nstar}, but for $\Lambda$ resonances.}
\label{lstar}
\end{figure}

The results are presented in Fig.~\ref{gsbar} for the ground state 
baryon octet with $J^P=1/2^+$ and the baryon decuplet with 
$J^P=3/2^+$. In Tables~\ref{nucdel} and \ref{strange} we show a 
comparison with all three and four star resonances. In our 
calculation we have assigned the $N(1440)$, $\Delta(1600)$, 
$\Sigma(1660)$ and $\Lambda(1600)$ resonances to the vibration 
characterized by $(v_1,v_2)=(1,0)$, and the $N(1710)$, 
$\Sigma(1940)$ and $\Lambda(1810)$ resonances to the 
$(v_1,v_2)=(0,1)$ vibration. The remaining resonances are assigned 
as rotational members of the ground band with $(v_1,v_2)=(0,0)$. 

We have followed the quark model 
assignments of Table~13.4 of \cite{PDG}, with the exception of 
the $\Sigma(1750)S_{11}$ resonance which we have assigned as 
$^{2}8_{1/2}[70,1^-]$, the lowest $S_{11}$ state with a 
mass of 1711 MeV. In our calculation the lowest four $S_{11}$ 
$\Sigma$ states occur at 1711, 1755, 1822 and 1974 MeV (for the 
assignments we refer to Tables~\ref{miss8} and \ref{miss10}; 
the second state belongs to the decuplet). 
In the nucleon and $\Lambda$ 
families the Roper resonance lies below the first excited negative 
parity resonance. We expect the same to be true for the $\Sigma$ 
hyperons. With our assignment, $\Sigma(1750)$ is the octet 
partner of $N(1535)$ and $\Lambda(1670)$, which is also supported 
by their $\eta$ decay properties \cite{Nefkens,Iachello}. 
In \cite{PDG} the two star $\Sigma(1620)$ resonance has been 
assigned as the $^{2}8_{1/2}[70,1^-]$ state, and the $\Sigma(1750)$ 
resonance instead as the $^{4}8_{1/2}[70,1^-]$ state. Our 
assignment of $\Sigma(1750)$ coincides with that of \cite{Rosner}. 
In the relatived quark model there are three low-lying $S_{11}$ 
$\Sigma$ resonances at 1630, 1675 and 1695 MeV \cite{rqm}. The 
first one was associated with the two star $\Sigma(1620)$ 
resonance, and the next one with the $\Sigma(1750)$ resonance. 

The $\Sigma(1940)D_{13}$ resonance was not assigned in Table~13.4 
of \cite{PDG}, whereas in \cite{Rosner} it was tentatively 
assigned as $^{4}8_{3/2}[70,1^-]$, the octet partner of $N(1700)$. 
In our calculation the lowest four $D_{13}$ $\Sigma$ states 
are the spin-orbit partners of the $S_{11}$ states at 1711, 1755, 
1822 and 1974 MeV, the first one of which has been associated 
with the $\Sigma(1670)$ resonance. We have assigned the $\Sigma(1940)$ 
resonance as a member of the $(v_1,v_2)=(0,1)$ vibrational band 
with $^{2}8_{3/2}[56,1^-]$ which occurs at 1974 MeV. This  
assignment is supported by its strong decay properties (see Sect. 6). 
In the relativized quark model there are three low-lying $D_{13}$ 
$\Sigma$ states at 1655, 1750 and 1755 MeV \cite{rqm}, of which 
the first two were associated with the $\Sigma(1670)$ and 
$\Sigma(1940)$ resonances. 

\subsection{Singlet resonances}

There are three states which show a deviation of about 100 MeV or 
more from the data: the $\Lambda^*(1405)$, $\Lambda^*(1520)$ 
and $\Lambda^*(2100)$ resonances are overpredicted by 236, 
121 and 97 MeV, respectively. These three resonances are 
assigned as singlet states in Table~\ref{strange} (and were 
not included in the fitting procedure). An additional energy 
shift for the singlet states (without effecting the masses of 
the octet and decuplet states) can be obtained by adding to 
the mass formula of Eq.~(\ref{massformula}) a term that only 
acts on the singlet states 
\ba
M^2 \;\rightarrow\; M^2 + 
\Delta M^2 \, \delta_{g_1,0} \delta_{g_2,0} ~.
\ea
This corresponds to a shift in the singlet masses of 
$M \sqrt{1+(\Delta M^2/M^2)} \approx M[1+(\Delta M^2/2M^2)]$. 
Since spin-orbit partners are shifted by the same amount, the mass 
splitting of 115 MeV between $\Lambda^*(1405)$ and $\Lambda^*(1520)$ 
cannot be reproduced by this mechanism. In principle, this 
splitting can be obtained from a spin-orbit interaction. However, 
in the rest of the baryon spectra there is no evidence  for such 
a large spin-orbit coupling. This problem is common 
to $qqq$ models of baryons ({\it e.g.} the constituent 
quark model with chromodynamics, either in its nonrelativistic 
\cite{nrqm} or its relativized form \cite{rqm}, and the chiral 
constituent quark model \cite{Plessas} all overpredict the 
$\Lambda^*(1405)$ mass). Another explanation for the mass splitting 
between $\Lambda^*(1520)$ and $\Lambda^*(1405)$ is the proximity of 
the $\Lambda^*(1405)$ resonance to the $N \overline{K}$ threshold. 
The inclusion of the coupling to the $N \overline{K}$ and 
$\Sigma \pi$ decay channels produces a downward shift of the $qqq$ 
state toward or even below the $N \overline{K}$ threshold 
\cite{Arima}. Such an interpretation is supported by the strong 
and electromagnetic couplings (see Sects. 6 and 8). 
In a chiral meson-baryon Lagrangian approach with an 
effective coupled-channel potential the $\Lambda^*(1405)$ 
resonance emerges as a quasi-bound state of $N \overline{K}$ 
\cite{Kaiser}. 

\subsection{Missing resonances} 

In Tables~\ref{nucdel} and \ref{strange} we presented the 
model states that could be associated with a three or four 
star resonance. In Tables~\ref{miss8}, \ref{miss10} and \ref{miss1} 
we show the masses of all low-lying octet, decuplet and singlet 
baryons. Since in the present approach no spin-orbit coupling 
has been taken into account, the states are grouped into multiplets 
labeled by $L$, $S$ and $|L-S| \leq J \leq L+S$. The multiplets 
for which at least one of its members has been associated with a  
three or four star resonance in Tables~\ref{nucdel} or \ref{strange} 
are labeled by $^{\dagger}$. Tentative assignments of one or two 
star resonances are indicated by $^{\ddagger}$. 
As in any $qqq$ model of baryons there 
are many more calculated states than have been observed. The lowest 
socalled `missing' resonances of the octet are associated 
with the $^{2}8_J[20,1^+]$ state. Their calculated mass is given 
by 1713, 1849, 1826 and 1957 MeV for the $N$, $\Sigma$, $\Lambda$ 
and $\Xi$ resonances. The search for these `missing' resonances 
is important in order to verify the assignments of the resonances 
and to distinguish between different models of baryons, such as 
three quark $qqq$ vs. quark-diquark $q-qq$ models which have 
less missing states because of the smaller number of 
degrees of freedom. 

In a recent three-channel multi-resonance 
amplitude analysis by the Zagreb group \cite{Zagreb} evidence was 
found for the existence of a third low-lying $P_{11}$ state 
at $1740 \pm 11$ MeV. The first two $P_{11}$ states 
at $1439 \pm 19$ MeV and $1729 \pm 16$ MeV correspond to the 
$N(1440)$ and $N(1710)$ resonances of the PDG \cite{PDG}. 
These $P_{11}$ states were associated with the states at 1540, 
1770 and 1880 MeV in the relativized quark model \cite{CLRS}. 
In the present calculation, they occur at 1444, 1683 and 1713 MeV, 
in good agreement with the analysis of the Zagreb group.  

A recent analysis of new data on kaon photoproduction \cite{Tran} 
has shown evidence for a $D_{13}$ resonance at 1895 MeV 
\cite{Mart}. In the present calculation, there are several 
possible assignments (see Table~\ref{miss8}). 
The lowest state that can be assigned to this new resonance 
is a vibrational excitation $(v_1,v_2)=(0,1)$ with 
$^{2}8_{3/2}[56,1^-]$ and mass 1847 MeV. This state belongs to 
the same vibrational band as the $N(1710)$ resonance, and is 
the octet partner of $\Sigma(1940)$. 
Another possible assignment is as a member of the ground state 
band $(v_1,v_2)=(0,0)$ with $^{2}8_{3/2}[70,2^-]$ and mass 1874 MeV. 
However, this state is completely decoupled in strong and 
electromagnetic decays. 
Finally, there is a state that belongs to the same vibrational 
band $(v_1,v_2)=(1,0)$ as the $N(1440)$ Roper resonance with 
$^{2}8_{3/2}[70,1^-]$ and mass 1909 MeV. 
As far as the mass is concerned all three assignments are 
possible. The strong couplings for these states provide a more 
sensitive tool to determine the most likely assignment 
(see Sect. 6). In the relativized quark model 
a $D_{13}$ state has been predicted at 1960 MeV \cite{rqm}. 


\section{Strong couplings}
\setcounter{equation}{0}

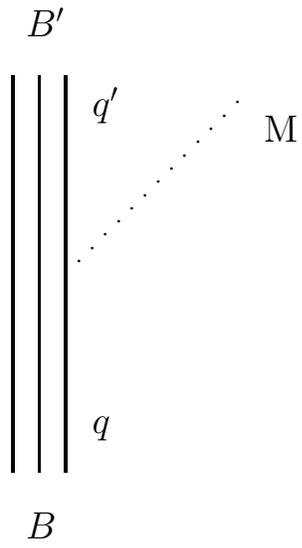
\begin{figure}
\centering
\Large
\setlength{\unitlength}{1.0pt}
\begin{picture}(200,250)(0,-25)
\thicklines
\put ( 25, 25) {\line (0,1){150}}
\put ( 35, 25) {\line (0,1){150}}
\put ( 45, 25) {\line (0,1){150}}
\put ( 30,  0) {$B$}
\put ( 30,190) {$B^{\prime}$}
\put ( 55, 40) {$q$}
\put ( 55,160) {$q^{\prime}$}
\multiput ( 45,100)( 5, 5){14}{\circle*{0.2}}
\put (120,150) {M}
\end{picture}
\normalsize
\vspace{0.5cm}
\caption[]{Elementary meson emission}
\label{qqM}
\end{figure}

Strong couplings provide an important test of baryon wave functions, 
and can be used to distinguish between different models of baryon 
structure. Here we consider strong decays of baryons by the emission 
of a pseudoscalar meson
\ba
B \rightarrow B^{\prime} + M ~.
\ea
Several forms have been suggested 
for the form of the operator inducing the strong transition 
\cite{LeYaouanc}. We use here the simple form \cite{KI}
\ba
{\cal H}_s &=& \frac{1}{(2\pi)^{3/2} (2k_0)^{1/2}}
\sum_{j=1}^{3} X^M_{j} \left[
2g \, (\vec{s}_j \cdot \vec{k}) \mbox{e}^{-i \vec{k} \cdot \vec{r}_j}
+ h \, \vec{s}_j \cdot
(\vec{p}_j \, \mbox{e}^{-i \vec{k} \cdot \vec{r}_j} +
\mbox{e}^{-i \vec{k} \cdot \vec{r}_j} \, \vec{p}_j) \right] ~, 
\label{hs}
\ea
where $\vec{r}_j$, $\vec{p}_j$ and $\vec{s}_j$ are the coordinate,
momentum and spin of the $j$-th constituent, respectively;
$k_0$ is the meson energy and $\vec{k}=k \hat z$ denotes the momentum
carried by the outgoing meson. The coefficients $g$ and $h$ denote the 
strength of the two terms in the transition operator of Eq.~(\ref{hs}).
The flavor operator $X^M_j$ corresponds to the emission of an
elementary meson by the $j$-th constituent:
$q_j \rightarrow q_j^{\prime} + M$ (see Figure~\ref{qqM}).

Using the symmetry of the wave functions, transforming to Jacobi
coordinates, integrating over the baryon center of mass coordinate, 
and adopting the rest frame of the initial baryon, the operator of 
Eq.~(\ref{hs}) reduces to \cite{strong} 
\ba
{\cal H}_s &=& \frac{1}{(2\pi)^{3/2} (2k_0)^{1/2}} \, 6 X^M_{3} 
\left[ (gk-\frac{1}{6}hk) \, s_{3,z} \hat U - h \, s_{3,z} \hat T_z 
- \frac{1}{2} h \, (s_{3,+} \hat T_- + s_{3,-} \hat T_+) \right] ~,
\label{hstrong}
\ea
with
\ba
\hat U &=& \mbox{e}^{i k \sqrt{\frac{2}{3}} \lambda_z} ~,
\nonumber\\
\hat T_m &=& \frac{1}{2} \left( \sqrt{\frac{2}{3}} \, p_{\lambda,m} \,
\mbox{e}^{i k \sqrt{\frac{2}{3}} \lambda_z} +
\mbox{e}^{i k \sqrt{\frac{2}{3}} \lambda_z} \,
\sqrt{\frac{2}{3}} \, p_{\lambda,m} \right) ~. \label{utop}
\ea
The calculation of the matrix elements of ${\cal H}_s$
can be done in configuration space ($\vec{\rho}$, $\vec{\lambda}$)
or in momentum space ($\vec{p}_{\rho}$, $\vec{p}_{\lambda}$). The mapping
onto the algebraic space of $U(7)$ is a convenient way to carry out the
calculations, much in the same way as the mapping of coordinates and
momenta onto creation and annihilation operators in the harmonic
oscillator space.
The operators $\hat U$ and $\hat T_m$ can be expressed algebraically 
by first making the replacement 
$\vec{p}_{\lambda}/m_3 \rightarrow -i k_0 \vec{\lambda}$ \cite{MB} 
and then mapping the coordinates onto the algebraic operators,
$\sqrt{2/3} \, \lambda_m \rightarrow \beta \hat D_{\lambda,m}/X_D$
\cite{BIL,emff,strong}. The result is 
\ba
\hat U &=& \mbox{e}^{i k \beta \hat D_{\lambda,z}/X_D} ~,
\nonumber\\
\hat T_m &=& - \frac{i m_3 k_0 \beta}{2X_D} \left(
\hat D_{\lambda,m} \, \mbox{e}^{i k \beta \hat D_{\lambda,z}/X_D} +
\mbox{e}^{i k \beta \hat D_{\lambda,z}/X_D} \,
\hat D_{\lambda,m} \right) ~. 
\label{algop}
\ea
The dipole operator $\hat D_{\lambda,m}$ is a generator of $U(7)$ and
$X_D$ is its normalization, as discussed in \cite{BIL,emff}. 
The spatial matrix elements of $\hat U$ and $\hat T_m$ are obtained 
in the collective model of baryons by folding with a distribution 
function $g(\beta)$ of charge and magnetization over the entire 
volume 
\ba
g(\beta) &=& \beta^2 \, \mbox{e}^{-\beta/a} \, /2a^3 ~.  \label{gbeta}
\ea
All spatial matrix elements can be expressed in terms of the 
collective form factors 
\ba
{\cal F}(k) &=& \int d \beta \, g(\beta) \,
\langle \psi' \, | \, \hat U \, | \, \psi \rangle ~,
\nonumber\\
{\cal G}_m(k) &=& \int d \beta \, g(\beta) \,
\langle \psi' \, | \, \hat T_m \, | \, \psi \rangle ~.
\label{fg}
\ea
Here $| \psi \rangle$ and $| \psi' \rangle$ denote the spatial 
wave functions of the initial and final baryons. 

For strong decays in which the initial baryon $B$ has angular
momentum $\vec{J}=\vec{L}+\vec{S}$ and in which the final baryon 
$B^{\prime}$ has $L^{\prime}=0$ and thus $J^{\prime}=S^{\prime}$, 
the helicity amplitudes in the collective model are then given by 
\ba
A_{\nu}(k) &=& \int d\beta \, g(\beta) \, 
\langle \Psi'(0,S^{\prime},S^{\prime},\nu)  
\, | \, {\cal H}_s \, | \, \Psi(L,S,J,\nu) \rangle ~, 
\ea
Here $| \Psi(L,S,J,M_J) \rangle$ and 
$| \Psi'(0,S',S',\nu) \rangle$ denote the (space-spin-flavor) 
angular momentum coupled wave functions of the initial and final 
baryons, respectively (see Appendix~B). 
The final baryon is a ground state baryon 
belonging either to the octet $^{2}8_{1/2}[56,0^+]_{(0,0);0}$ or 
the decuplet $^{4}10_{3/2}[56,0^+]_{(0,0);0}$. 
The helicity amplitudes can be expressed in terms of a spatial 
matrix element and a spin-flavor matrix element 
\ba
A_{\nu}(k) &=& \frac{1}{(2\pi)^{3/2} (2k_0)^{1/2}} \left[
\langle L, 0,S,\nu   | J,\nu \rangle \, \zeta_0 Z_0(k) + \frac{1}{2}
\langle L, 1,S,\nu-1 | J,\nu \rangle \, \zeta_+ Z_-(k) \right.
\nonumber\\
&& + \left. \frac{1}{2}
\langle L,-1,S,\nu+1 | J,\nu \rangle \, \zeta_- Z_+(k) \right] ~,
\label{anu}
\ea
with 
\ba
Z_0(k) &=& 6 \, (gk - \frac{1}{6} hk) \, {\cal F}(k)
- 6h \, {\cal G}_z(k) ~,
\nonumber\\
Z_{\pm}(k) &=& -6h \, {\cal G}_{\pm}(k) ~. \label{radme}
\ea
In Table~\ref{collff} we present the collective form factor 
${\cal F}(k)$ (for details of the derivation we refer to \cite{BIL}). 
The form factors ${\cal G}_m(k)$ are given by 
\ba
{\cal G}_z(k) &=& - \delta_{M,0} \, m_3 k_0 a \, 
\frac{\mbox{d} {\cal F}(k)}{\mbox{d} ka} ~, 
\nonumber\\
{\cal G}_{\pm}(k) &=& \mp \delta_{M,\pm 1} \, m_3 k_0 a \, 
\sqrt{L(L+1)} \, \frac{{\cal F}(k)}{ka} ~. 
\label{gff} 
\ea
For any other model of baryons with the same spin-flavor structure, 
the corresponding results can be obtained by replacing Table~\ref{collff} 
with the appropriate table (for example, by using harmonic oscillator wave 
functions as discussed in \cite{BIL}).

The coefficients $\zeta_m$ are the spin-flavor matrix elements of 
$X^M_{3} \, s_{3,m}$ which can either be evaluated for each channel 
separately, or more conveniently, by using the Wigner-Eckart theorem 
and isoscalar factors of $SU_{\rm f}(3)$ \cite{deSwart}. 
The flavor wave functions are labeled by the quantum numbers $(p,q),I,Y$ 
corresponding to the reduction 
$SU_{\rm f}(3) \supset SU_{\rm I}(2) \otimes U_{\rm Y}(1)$. 
In this notation $(p,q)=(g_1-g_2,g_2)$, and hence we have 
$(p,q)=(1,1)$, $(3,0)$ or $(0,0)$ for the baryon flavor octet, 
decuplet and singlet, respectively, and $(p,q)=(1,1)$ or $(0,0)$ for the meson
flavor octet and singlet, respectively. 
The spin-flavor matrix elements $\zeta_m$ for a given isospin channel 
can be expressed as 
\ba 
\zeta_m &=& \sum_{\gamma} \left< \left. \begin{array}{cc}
(p_f,q_f) & (p,q) \\ I_f,Y_f & I,Y \end{array}
\right| \begin{array}{c} (p_i,q_i)_{\gamma} \\ I_i,Y_i \end{array} \right> 
\alpha_{m,\gamma} ~. 
\label{alfa}
\ea
The sum over $\gamma$ is over different multiplicities. 
The $SU(3)$ isoscalar factor which appears in Eq.~(\ref{alfa}) 
depends on the flavor multiplets $(p,q)$, the isospin $I$ and the 
hypercharge $Y$. A compilation of the $SU(3)$ isoscalar factors relevant 
for strong decays of baryons can be found in \cite{PDG}. 
Results for a specific charge channel can be obtained by multiplying
$\zeta_m$ with the appropriate isospin Clebsch-Gordan coefficient 
$\langle I_f,M_{I_f},I,M_{I}|I_i,M_{I_i} \rangle$. 

In Tables~\ref{B8M8}--\ref{B10M1} we present the coefficients 
$\alpha_{m,\gamma}$ for strong decays into octet or decuplet 
baryons emitting a pseudoscalar meson (either octet or singlet). 
For strong decays of nonstrange baryons into the $N\pi$, $N\eta$, 
$\Delta\pi$ and $\Delta\eta$ channels the coefficients $\zeta_m$ 
are given explicitly in Tables~III and IV of \cite{strong}.
Inspection of Tables~\ref{B8M8}--\ref{B10M1} and 
the isoscalar factors on page 184 of \cite{PDG} yields some 
interesting selection rules: 
(i) the $B_{10} \rightarrow B_{10} + M_8$ decay 
\ba
\Sigma^* &\rightarrow& \Sigma^* + \eta_8 ~,
\ea
is forbidden since the $SU(3)$ isoscalar factor vanishes, and 
(ii) there is a spin-flavor selection rule for the 
$^{4}8[70,L^P] \rightarrow \, ^{2}8[56,0^+] + M_8$ decays 
\ba
N &\rightarrow& \Lambda + K ~, 
\nonumber\\
\Lambda &\rightarrow& N + \overline{K} ~, 
\nonumber\\
\Xi &\rightarrow& \Xi + \eta_8 ~, 
\ea
which is similar to the Moorhouse selection rule in electromagnetic 
couplings \cite{Moorhouse}. 
However, the octet $\eta_8$ and singlet $\eta_1$ may mix because of 
$SU_{\rm f}(3)$ flavor symmetry breaking. The physical mesons 
$\eta$ and $\eta^{\prime}$ are then given in terms of a mixing angle 
\ba
\eta &=& \eta_8 \, \cos \theta_P - \eta_1 \, \sin \theta_P ~,
\nonumber\\
\eta^{\prime} &=& \eta_8 \, \sin \theta_P + \eta_1 \, \cos \theta_P ~.
\label{etamix}
\ea
The above mentioned forbidden two-body decays into a baryon and 
the octet meson $\eta_8$, are allowed for the physical mesons 
$\eta$ and $\eta^{\prime}$ via the octet-singlet mixing. 


\section{Comparison with experimental strong decays}
\setcounter{equation}{0}

With the definition of the transition operator in Eq.~(\ref{hs}) 
and the helicity amplitudes, the decay widths for a
specific isospin channel are given by \cite{LeYaouanc}
\ba
\Gamma(B \rightarrow B^{\prime} + M) &=& 2 \pi \rho \,
\frac{2}{2J+1} \sum_{\nu>0} | A_{\nu}(k) |^2 ~. 
\label{dw}
\ea
Here we adopt the procedure of \cite{LeYaouanc}, in which the 
decay widths are calculated in the rest frame of the decaying 
resonance, and in which the relativistic expression for the phase 
space factor $\rho$ as well as for the momentum $k$ of the emitted 
meson are retained. The expressions for $k$ and $\rho$ are 
\ba
k^2 &=& -m_M^2 + \frac{(m_B^2-m_{B^{\prime}}^2+m_M^2)^2}{4m_B^2} ~,
\nonumber\\
\rho &=& 4 \pi \frac{E_{B^{\prime}} E_M k}{m_B}
\ea
with $E_{B^{\prime}}=\sqrt{m_{B^{\prime}}^2+k^2}$ and
$E_{M}=\sqrt{m_{M}^2+k^2}$.
We consider here the strong decays of baryons in which a 
pseudoscalar meson (either octet or singlet) is 
emitted. The present calculation is an extension of \cite{strong} 
in which we only discussed nonstrange decays of nonstrange baryons. 

The calculated widths depend on the two parameters $g$ and $h$ 
in the transition operator of Eq.~(\ref{hstrong}), and on the scale 
parameter $a$ of Eq.~(\ref{gbeta}). In accordance with \cite{strong} 
we keep $g$, $h$ and $a$ fixed for {\em all} resonances and 
{\em all} decay channels with the values $g=1.164$ fm, $h=-0.094$ 
fm and $a=0.232$ fm (we note here that in \cite{strong} the values 
of $g$ and $h$ were given in GeV$^{-1}$ instead of in fm). 
The decay widths of resonances that have been interpreted as 
a vibrational excitation of the string configuration of 
Fig.~\ref{geometry} depend on the coefficient $\chi_1$ 
for $(v_1,v_2)=(1,0)$ (the $N(1440)$, $\Delta(1600)$, 
$\Sigma(1660)$ and $\Lambda(1600)$ resonances), or $\chi_2$ for 
$(v_1,v_2)=(0,1)$ (the $N(1710)$, $\Sigma(1940)$ and 
$\Lambda(1810)$ resonances) (see Table~\ref{collff}). 
These coefficients are proportional 
to the intrinsic matrix element for each type of vibration. 
Here they are taken as constants with the values $\chi_1=1.0$ 
\cite{Tomasi} and $\chi_2=0.7$. 
For the pseudoscalar $\eta$ mesons we introduce a mixing angle 
$\theta_P=-23^{\circ}$ between the octet and singlet mesons 
\cite{PDG}. This value is consistent with 
that determined in a study of meson spectroscopy \cite{meson}. 

In comparison with other studies, we note that in the 
calculation in the nonrelativistic quark model of \cite{KI} 
an elementary emission model is used, just as in the present 
calculation, but with the difference that the decay widths are 
parametrized by four reduced partial wave amplitudes instead of 
the two elementary amplitudes $g$ and $h$. Furthermore, the 
momentum dependence of these reduced amplitudes is represented 
by a constant. The calculations in the relativized quark model 
are done in a pair-creation model for the decay and involve a 
different assumption on the phase space factor \cite{CR}. 
Both the nonrelativistic and relativized quark model calculations
include the effects of mixing induced by the hyperfine interaction,
which in the present calculations are not taken into account. 
It is important to note that we present a comparison of decay 
widths, rather than of decay amplitudes as was done in \cite{KI} 
and \cite{CR} for the nonrelativistic and relativized quark models.  

In Tables~\ref{nuc}--\ref{xi} we compare the experimental 
strong decay widths of three and four star baryon resonances from 
the most recent compilation by the Particle Data Group \cite{PDG} 
with the results of our calculation for the nucleon, $\Delta$, 
$\Sigma$, $\Lambda$ and $\Xi$ families. We have used the 
experimental value of the mass of the decaying baryon. 
Strong couplings of missing resonances belonging to the 
flavor octet, decuplet and singlet are presented in 
Tables~\ref{strong1}--\ref{strong4}, \ref{strong5}--\ref{strong8} 
and \ref{strong9}, respectively. 

The calculated decay widths are to a large extent a consequence of 
spin-flavor symmetry and phase space. The use of the collective form 
factors of Table~\ref{collff} introduces a power-law dependence on 
the meson momentum $k$, compared to, for example, an exponential 
dependence for harmonic oscillator form factors. Our results for 
the strong decay widths are in fair overall agreement with the 
available data, and show that the combination of a collective 
string-like $qqq$ model of baryons and a simple 
elementary emission model for the decays can account for the 
main features of the data. There are a few exceptions which could 
indicate evidence for the importance of degrees of freedom which 
are outside the present $qqq$ model of baryons. 

\subsection{Nucleon resonances}

The $\pi$ and $\eta$ decays of nucleon resonances have already 
been discussed in \cite{strong}. Whereas the $\pi$ decays are in 
fair agreement with the data, the $\eta$ decays of octet baryons 
show an unusual pattern: the $S$-wave states $N(1535)$, 
$\Sigma(1750)$ and $\Lambda(1670)$ all are found experimentally 
to have a large branching ratio to the $\eta$ channel, 
whereas the corresponding phase space factor is very small 
\cite{Nefkens}. The small calculated $\eta$ widths ($<0.5$ MeV) 
for these resonances are due to a combination of spin-flavor 
symmetry and the size of the phase space factor. The results of 
our analysis suggest that the observed $\eta$ widths are not due 
to a conventional $qqq$ state, but may rather indicate evidence 
for the presence of a state in the same mass region of a more 
exotic nature, such as a pentaquark configuration 
$qqqq\overline{q}$ or a quasi-molecular $S$-wave resonance 
$qqq$-$q\overline{q}$ just below or above threshold, bound by Van 
der Waals type forces (for example $N\eta$, $\Sigma K$ 
or $\Lambda K$ \cite{Kaiser}). In order to answer this question 
one has to carry out an analysis, similar to the present one, 
of the other configurations. 

The $K$ decays are suppressed with respect to the $\pi$ 
decays because of phase space. In addition, the decay of the 
$N(1650)$, $N(1675)$ and $N(1700)$ resonances into 
$\Lambda K$ is forbidden by the spin-flavor selection 
rule for the decay of $^{4}8[70,L^P]$ nucleon states into 
this channel (see Sect. 5). For $N(1675)$ and $N(1700)$ 
only an upper limit is known, whereas the $N(1650)$ resonance 
has an observed width of $12 \pm 7$ MeV. However, this resonance 
is just above the $\Lambda K$ threshold which may lead to a 
coupling to a quasi-bound meson-baryon $S$ wave resonance. 
A study of the effect of spin-flavor symmetry breaking on these 
decays is in progress. 

\subsection{Delta resonances}

The strong decay widths of the $\Delta$ resonances are in very 
good agreement with the available experimental data. The same 
holds for the other resonances that have been assigned as 
decuplet baryons: $\Sigma^*(1385)$, $\Sigma^*(2030)$ and 
$\Xi^*(1530)$. For the decuplet baryons there is no $S$ state 
around the threshold of the various decay channels, so therefore 
there cannot be any coupling to quasi-molecular configurations. 
Just as for the nucleon resonances, the $\eta$ and $K$ decays 
are suppressed by phase space factors. 

\subsection{Sigma resonances}

Strange resonances decay predominantly into the $\pi$ and 
$\overline{K}$ channel. Phase space factors suppress the $\eta$ 
and $K$ decays. The main discrepancy is found 
for $\Sigma(1750)$. In the discussion of nucleon resonances 
it was suggested that the $S$ wave state $\Sigma(1750)$ is the 
octet partner of $N(1535)$. It has a large observed $\eta$ width 
despite the fact that there is hardly any phase space available 
for this decay. This may indicate that it has a large 
quasi-molecular component. 

The assignment of $\Sigma(1940)D_{13}$ as a member of the 
$(v_1,v_2)=(0,1)$ vibrational band with $^{2}8_{3/2}[56,1^-]$ 
and mass 1974 MeV is based on both its mass and its 
strong decay properties. If calculated with the observed mass, 
the other possible states, 
$^{2}10_{3/2}[70,1^-]$ and $^{4}8_{3/2}[70,1^-]$, 
both have very large widths ($\sim$ 100 MeV) in the 
$\Delta \overline{K}$ and $\Sigma^* \pi$ channels, which 
is not supported by the data. 

\subsection{Lambda resonances}

Also for $\Lambda$ resonances the $\eta$ and $K$ decays 
are suppressed with respect to the $\pi$ and $\overline{K}$ 
channels because of phase space factors. 
Table~\ref{lam} shows that the strong decays of $\Lambda$ 
resonances show more discrepancies with the data than the other 
families of resonances. 

We have assigned $\Lambda(1670)$ as the octet partner of 
$N(1535)$ and $\Sigma(1750)$. Its decay properties into the 
$\eta$ channel have already been discussed in Sect. 6.1 
on the nucleon resonances. 

The spin-flavor selection rule that 
was discussed in Sect. 5 forbids the decay of $^{4}8[70,L^P]$ 
$\Lambda$ states into the $N \overline{K}$ channel. Therefore, 
the calculated $N \overline{K}$ widths of $\Lambda(1800)$, 
$\Lambda(1830)$ and $\Lambda(2110)$ vanish, whereas all of 
them have been observed experimentally \cite{PDG}. 
The $\Lambda(1800)S_{01}$ state has large decay width into 
$N \overline{K}^*(892)$ \cite{PDG}. Since the mass of 
the resonance is just around the threshold of this channel, 
this could indicate a coupling with a quasi-molecular $S$ 
wave. The $N \overline{K}$ width of $\Lambda(1830)$ is 
relatively small ($6 \pm 3$ MeV), and hence in qualitative 
agreement with the selection rule. The situation for the 
the $\Lambda(2110)$ resonance is unclear. 

The $\Lambda^*(1405)$ resonance has a anomalously large decay 
width ($50 \pm 2$ MeV) into $\Sigma \pi$. This feature 
emphasizes the quasi-molecular nature of $\Lambda^*(1405)S_{01}$ 
due to the proximity of the $N \overline{K}$ threshold. 
It has been shown \cite{Arima} that the inclusion of the 
coupling to the $N \overline{K}$ and $\Sigma \pi$ decay 
channels produces a downward shift of the $qqq$ 
state toward or even below the $N \overline{K}$ threshold. 
In a chiral meson-baryon Lagrangian approach with an 
effective coupled-channel potential the $\Lambda^*(1405)$ 
resonance emerges as a quasi-bound state of $N \overline{K}$ 
\cite{Kaiser}. 

\subsection{Xi resonances} 

There is little experimental information available for the strong 
decays of baryons with strangeness $-2$ ($\Xi$) and $-3$ ($\Omega$). 
We find good agreement with the observed decay widths of the 
$\Xi(1820)$ (octet) and the $\Xi(1530)$ (decuplet) resonances 
(see Table~\ref{xi}). 

\subsection{Missing resonances}

For possible use in the analysis of new experimental data 
and in the search for missing resonances, we present in 
Tables~\ref{strong1}--\ref{strong9} the strong decay widths of 
the missing resonances of Tables~\ref{miss8}--\ref{miss1}. 
The states with $L^P=1^+$, $L^P=2^-$ and $[20,L^P]$ are decoupled 
because of spin-flavor symmetry. Most of the strong couplings of 
the low-lying missing resonances are small, which to a large 
extent explains their status \cite{missing}. Generally speaking, 
the orbital configurations that have the smallest strong 
couplings both for the octet, decuplet and singlet 
resonances are $[56,2^+]$ $(v_1,v_2)=(0,0)$, $[70,1^-](1,0)$ 
and $[70,1^-](0,1)$. It is interesting to note that in all 
cases the resonances associated with the configuration 
$[56,1^-](0,1)$ exhibit large decay widths. The only resonance 
that we have assigned as one of these is $\Sigma(1940)$. 
The majority of the missing resonances with sizeable decay 
widths belong either to the configuration $[56,1^-](0,1)$ or 
to $[70,2^+](0,0)$. 

Tables~\ref{strong1} and \ref{strong5} show that the 
missing nucleon and $\Delta$ resonances that are 
associated with the configurations $[70,2^+](0,0)$, 
$[70,0^+](0,1)$ and $[56,1^-](0,1)$ are predicted to have 
a large decay width into the $\pi$ channel. The $\eta$ and 
$K$ decays are suppressed with respect to the $\pi$ decays  
because of phase space. 
Strange baryons decay predominantly into the $\pi$ and 
$\overline{K}$ channels. The $\eta$ and $K$ widths are small in 
comparison, due to the available phase space. 
Inspection of Tables~\ref{strong2}--\ref{strong4} shows that 
the dominant decay channels of the missing octet baryons are 
$N \overline{K}$, $\Sigma \pi$, $\Delta \overline{K}$ for 
$\Sigma$ resonances, $N \overline{K}$, $\Sigma^* \pi$ for 
$\Lambda$ resonances, and $\Sigma \overline{K}$, $\Xi \pi$ for 
$\Xi$ resonances. Similarly, we see from 
Tables~\ref{strong6}--\ref{strong8} that the missing decuplet 
baryons are most likely to couple to $\Lambda \pi$, 
$\Sigma^* \pi$ for $\Sigma^*$ resonances, to 
$\Sigma \overline{K}$, $\Lambda \overline{K}$, $\Xi \pi$, 
$\Sigma^* \overline{K}$ for $\Xi^*$ resonances, and to 
$\Xi \overline{K}$ for $\Omega$ resonances. Finally, 
Table~\ref{strong9} shows that 
missing singlet baryons are most likely to decay into 
$\Lambda^* \rightarrow N \overline{K}$, $\Sigma \pi$. 


\section{Electromagnetic couplings}
\setcounter{equation}{0}

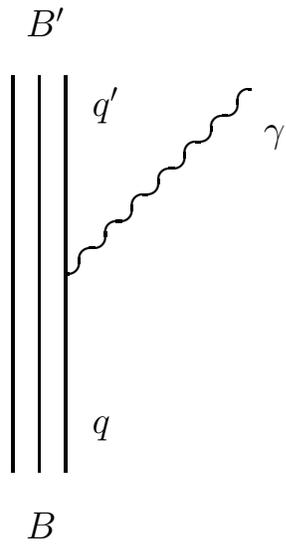
\begin{figure}
\centering
\Large
\setlength{\unitlength}{1.0pt}
\begin{picture}(200,250)(0,-25)
\thicklines
\put ( 25, 25) {\line (0,1){150}}
\put ( 35, 25) {\line (0,1){150}}
\put ( 45, 25) {\line (0,1){150}}
\put ( 30,  0) {$B$}
\put ( 30,190) {$B^{\prime}$}
\put ( 55, 40) {$q$}
\put ( 55,160) {$q^{\prime}$}
\multiput ( 45,105)(10,10){7}{\oval(10,10)[br]}
\multiput ( 55,105)(10,10){7}{\oval(10,10)[tl]}
\put (120,150) {$\gamma$}
\end{picture}
\normalsize
\vspace{0.5cm}
\caption[]{Photon emission}
\label{qqF}
\end{figure}

In constituent models, electromagnetic couplings arise from the 
coupling of the (point-like) constituent parts to the electromagnetic 
field \cite{copley}. We discuss here the 
case of the emission of a lefthanded photon 
\ba
B \rightarrow B' + \gamma ~, 
\ea
for which the nonrelativistic part of the transverse 
electromagnetic coupling is given by
\ba
{\cal H}_{em} &=& 2 \sqrt{\frac{\pi}{k_0}} \sum_{j=1}^{3} \mu_j e_j \, 
\left[ k s_{j,-} \, \mbox{e}^{-i \vec{k} \cdot \vec{r}_j}
+ \frac{1}{2g_j} (p_{j,-} \, \mbox{e}^{-i \vec{k} \cdot \vec{r}_j} +
\mbox{e}^{-i \vec{k} \cdot \vec{r}_j} \, p_{j,-}) \right] ~, 
\ea
where $\vec{r}_j$, $\vec{p}_j$ and $\vec{s}_j$ are the coordinate,
momentum and spin of the $j$-th constituent, respectively;
$k_0$ is the photon energy, and $\vec{k}=k \hat z$ 
denotes the momentum carried by the outgoing photon.  
The photon is emitted by the $j$-th constituent: 
$q_j \rightarrow q_j^{\prime} + \gamma$ (see Figure~\ref{qqF}).
The transition operator can be simplified by 
using the symmetry of the baryon wave functions, transforming to 
Jacobi coordinates and integrating over the baryon 
center-of-mass coordinate, to obtain 
\ba
{\cal H}_{em} &=& 6 \sqrt{\frac{\pi}{k_0}} \mu_3 e_3 \,  
\left[ k s_{3,-} \hat U - \frac{1}{g_3} \hat T_{-} \right] ~. 
\label{hem}
\ea
The operators $\hat U$ and $\hat T_-$ are given in Eq.~(\ref{algop}). 

The transverse helicity amplitudes between the final ground state 
baryon belonging either to the $J^P=1/2^+$ octet with 
$^{2}8_{1/2}[56,0^+]_{(0,0);0}$ or to the $J^P=3/2^+$ 
decuplet with $^{4}10_{3/2}[56,0^+]_{(0,0);0}$, 
and the initial (excited) state of a baryon resonance 
are expressed as \cite{BIL}
\ba
A_{\nu}(k) &=& \int d\beta \, g(\beta) \, 
\langle \Psi'(0,S^{\prime},S^{\prime},\nu-1) ) 
\, | \, {\cal H}_{em} \, | \, \Psi(L,S,J,\nu) \rangle ~, 
\nonumber\\
&=& 6 \sqrt{\frac{\pi}{k_0}} \, \Bigl [ \, k
\langle L,0;S,\nu   | J,\nu \rangle \, {\cal B}_{\nu} -
\langle L,1;S,\nu-1 | J,\nu \rangle \, {\cal A}_{\nu} \, \Bigr ] ~,
\label{helamp}
\ea
where $\nu=1/2$, $3/2$ indicates the helicity.
The orbit- and spin-flip amplitudes (${\cal A}_{\nu}$ and ${\cal B}_{\nu}$,
respectively) are given by
\ba
{\cal B}_{\nu} &=& \int \mbox{d} \beta \, g(\beta)\,
\langle \Psi'(0,0;S^{\prime},\nu-1) \, | \, \mu_3 \,
e_3 \, s_{3,-} \, \hat U \, | \, \Psi(L,0;S,\nu) \rangle ~,
\nonumber\\
{\cal A}_{\nu} &=& \int \mbox{d} \beta \, g(\beta)\,
 \langle \Psi'(0,0;S^{\prime},\nu-1) \, | \, \mu_3 \,
e_3 \, \hat T_{-}/g_3 \, | \, \Psi(L,1;S,\nu-1) \rangle ~.
\label{ab}
\ea
Here $| \Psi(L,M_L;S,M_S) \rangle$ denote the (space-spin-flavor) 
angular momentum uncoupled wave functions of the initial and final 
baryons (see Appendix~B). 

In Tables~\ref{gamma1} and \ref{gamma2} we show the orbit- and 
spin-flip amplitudes for some radiative hyperon decays. These 
results were obtained under the assumption of $SU_{\rm f}(3)$ 
flavor symmetry, {\it i.e.} $\mu_3=\mu_p$ and 
$g_3=g$. The following selection rules apply: 
(i) the $^{4}10[56] \rightarrow \,^{2}8[56] + \gamma$ decays 
\ba
\Sigma^{*,-} &\rightarrow& \Sigma^- + \gamma ~,
\nonumber\\
\Xi^{*,-} &\rightarrow& \Xi^- + \gamma ~,
\ea
are forbidden by $U$-spin conservation \cite{Landsberg}. These 
decays can only occur if flavor symmetry is broken ($m_d \neq m_s$). 
Also the $^{2}1[70] \rightarrow \,^{4}10[56] + \gamma$ decay 
\ba 
\Lambda^* &\rightarrow& \Sigma^{*,0} + \gamma ~, 
\ea
is forbidden by $U$-spin selection rules but, contrary 
to the previous cases, remains forbidden in the case of flavor 
symmetry breaking. 


\section{Comparison with experimental electromagnetic couplings} 
\setcounter{equation}{0}

Radiative hyperon decay widths can be calculated from the 
helicity amplitudes as \cite{LeYaouanc} 
\ba
\Gamma(B \rightarrow B^{\prime} + \gamma) &=& 2 \pi \rho \, 
\frac{1}{(2\pi)^3} \, \frac{2}{2J+1} \sum_{\nu>0} | A_{\nu}(k) |^2 ~. 
\label{gw}
\ea
Just as for the strong couplings, the electromagnetic decay 
widths are calculated assuming $SU_{\rm sf}(6)$ spin-flavor 
symmetry and using the rest frame of the decaying resonance  
\ba
k &=& \frac{m_B^2-m_{B^{\prime}}^2}{2m_B} ~,
\nonumber\\
\rho &=& 4 \pi \frac{E_{B^{\prime}} k^2}{m_B}
\ea
with $E_{B^{\prime}}=\sqrt{m_{B^{\prime}}^2+k^2}$. The 
scale parameter $a$ was determined in a simultaneous fit to 
the proton charge radius, the proton electric and magnetic form 
factors and the neutron magnetic form factor to be $a = 0.232$ fm 
\cite{emff}. This is the same value as has been determined 
independently in a fit of the $N \pi$ decay widths of 
nucleon and delta resonances \cite{strong}. 
For all cases we take the quark $g$ factors $g=1$. 
The quark scale magnetic moment is equal to the proton magnetic 
moment $\mu_p$, which corresponds to a 
constituent quark mass $m=0.336$ GeV. 

Recently, the SELEX collaboration has 
measured the charge radius of the $\Sigma^-$ hyperon. The 
preliminary value is $\langle r^2 \rangle_{\Sigma^-} = 
0.60 \pm 0.08 \pm 0.08$ fm$^2$ \cite{b98}. This value 
is in good agreement with our predicted value of 
$\langle r^2 \rangle_{\Sigma^-} = \langle r^2 \rangle_p 
= 12 a^2 = 0.65$ fm$^2$. 

The radiative decays between baryons with $L=0$ and 
$L^{\prime}=0$ only involve the magnetic transitions. 
The corresponding widths can be expressed in terms of the 
transition magnetic moments $\mu_{BB'}(k)$ via 
\ba 
\sum_{\nu>0} | A_{\nu}(k) |^2 &=& 4 \pi k \, \mu_{BB'}^2(k) ~. 
\ea
In Table~\ref{trmm} we show the transition moments for the 
$^{2}8[56] \, \rightarrow \,^{2}8[56] + \gamma$ and 
$^{4}10[56] \, \rightarrow \,^{2}8[56] + \gamma$ transitions. 
In the absence of the form factor ({\it i.e.} ${\cal F}(k)=1$),  
we recover the symmetry relations between the decuplet 
to octet transitions \cite{Pais,Landsberg}. For the conventions 
used in Appendices A and B we obtain the relations 
\ba
\mu_{\Sigma^0 \Lambda} &=& \frac{1}{\sqrt{3}} \mu_p ~,
\nonumber\\
\mu_{\Delta^+ p} &=& \mu_{\Delta^0 n} \;=\; 
-\mu_{\Sigma^{*,+} \Sigma^+} \;=\; 
-2\mu_{\Sigma^{*,0} \Sigma^0} 
\nonumber\\
&=& \frac{2}{\sqrt{3}} \mu_{\Sigma^{*,0} \Lambda} \;=\;  
-\mu_{\Xi^{*,0} \Xi^0} \;=\; \frac{2\sqrt{2}}{3} \mu_p ~, 
\nonumber\\
\mu_{\Sigma^{*,-} \Sigma^-} &=&  
\mu_{\Xi^{*,-} \Xi^-} \;=\; 0 ~.
\ea
The numerical values are given in the third column of 
Table~\ref{trmm}. A comparison with the last column shows the 
reduction of the transition magnetic moments due to the form 
factor ${\cal F}(k)=1/(1+k^2 a^2)^2$. 

The experimental information on radiative decays of hyperons  
is very limited. In Table~\ref{key} we present the radiative 
decay widths of low-lying hyperon resonances, and compare 
wherever possible with the data. The $\Delta^+ \rightarrow 
p + \gamma$ decay width is underpredicted by 35 $\%$, a common 
feature of all $qqq$ constituent quark models. This discrepancy 
has been shown to be due to nonresonant meson-exchange 
mechanisms \cite{tshlee}. Just as for the energies and 
the strong decays, the $\Lambda^*(1405)$ resonance shows large 
deviations for the radiative decay widths, which once agains 
confirms its uncertain nature as a $qqq$ state. The forbidden 
decays $\Lambda^*(1405) \rightarrow \Sigma^{*,0} + \gamma$ and 
$\Lambda^*(1520) \rightarrow \Sigma^{*,0} + \gamma$ have not been 
observed. For comparison we also present the radiative decay widths 
of decuplet hyperons as obtained from lattice calculations 
\cite{lattice} and from a chiral constituent quark model with 
electromagnetic exchange currents between quarks \cite{Wagner}. 
The negative parity hyperon decay widths for 
$\Sigma^{*,-} \rightarrow \Sigma^- + \gamma$ and 
$\Xi^{*,-} \rightarrow \Xi^- + \gamma$ have a small nonvanishing 
value in \cite{lattice,Wagner}, whereas in the present 
calculation they are forbidden by flavor symmetry 
selection rules. In a subsequent publication we plan to investigate 
the effects of $SU_{\rm f}(3)$ flavor symmetry breaking on the 
electromagnetic couplings. 


\section{Summary and conclusions}
\setcounter{equation}{0}

We have presented in this article a systematic analysis of spectra 
and transition rates of strange baryons in the framework of a 
collective string-like $qqq$ model, in which the orbitally 
excited baryons are interpreted as collective rotations and 
vibrations of the strings. The algebraic structure of the model, 
both for the internal degrees of freedom of spin-flavor-color 
and for the spatial degrees of freedom, has been used to derive 
transparent results, such as a mass formula, selection rules and 
closed expressions for strong and electromagnetic couplings. 

The situation is similar to that encountered for nonstrange baryons. 
While spectra are reasonably well described, transition rates, 
especially strong decay widths are only qualitatively described. 
The combination of a collective string-like $qqq$ model of baryons 
and a simple elementary emission model for the decays can account 
for the main features of the data. 
The main discrepancies are found for the low-lying $S$-wave states,  
specifically $N(1535)$, $\Sigma(1750)$, $\Lambda^*(1405)$, 
$\Lambda(1670)$ and $\Lambda(1800)$. All of these resonances 
have masses which are close to the threshold of a meson-baryon 
decay channel, and hence they could mix with a quasi-molecular 
$S$ wave resonance of the form $qqq-q\overline{q}$. 
In contrary, decuplet baryons have no low-lying $S$  
states with masses close to the threshold of a particular decay 
channel, and their spectroscopy is described very well. 
The results of our analysis suggest that in future experiments 
particular attention be paid to the resonances mentioned above 
in order to elucidate their structure, and to look for evidence 
of the existence of exotic (non $qqq$) configurations of quarks 
and gluons.  

In our calculations we have included only a diagonal breaking of the 
spin-flavor symmetry. This seems to be a good approximation to the
actual situation and no major discrepancy appear to be related to
non-diagonal breakings. A study of the effects of $SU_{\rm f}(3)$ 
flavor symmetry breaking due to different quark masses on the 
radiative decays and strong couplings is in progress, and will be 
published separately. 

This paper concludes our analysis of $q^{3}$ configurations in 
baryons. The next step is the study of more complex configurations 
of quarks and gluons, such as hybrid quark-gluon states 
$qqq-g$, pentaquark states $q^{4}\overline{q}$ and multiquark 
meson-baryon bound states $qqq-q\overline{q}$. 

\section*{Acknowledgements}

This work is supported in part by DGAPA-UNAM under 
project IN101997, by CONACyT under project 32416-E 
and by D.O.E. Grant DE-FG02-91ER40608.


\appendix
\section{Spin-flavor wave functions}
\setcounter{equation}{0}

Here we list the conventions used for the spin and flavor 
wave functions which are consistent with the choice of 
Jacobi coordinates of Eq.~(\ref{jacobi}). They coincide 
with the conventions of \cite{KI}. 

\subsection{Spin wave functions} 

The spin wave functions $|S,M_S\rangle$ are given by \cite{KI}:
\ba
|1/2,1/2\rangle &:& \chi_{\rho} \;=\; 
[ |\uparrow \downarrow \uparrow \, \rangle 
- |\downarrow \uparrow \uparrow \, \rangle ]/\sqrt{2} ~, 
\nonumber\\ 
&:& \chi_{\lambda} \;=\; 
[2|\uparrow \uparrow \downarrow \, \rangle 
- |\uparrow \downarrow \uparrow \, \rangle 
- |\downarrow \uparrow \uparrow \, \rangle ]/\sqrt{6} ~, 
\nonumber\\ 
|3/2,3/2\rangle &:& \chi_S \;=\; 
|\uparrow \uparrow \uparrow \, \rangle ~.
\ea
We only show the state with the largest component of the projection
$M_S=S$. The other states are obtained by applying the lowering operator 
in spin space.

\subsection{Flavor wave functions}

For the flavor wave functions $|(p,q),I,M_I,Y \rangle$ we adopt the 
convention of \cite{deSwart} with $(p,q)=(g_1-g_2,g_2)$.\\
(i) The octet baryons $(p,q)=(1,1)$: 
\ba
|(1,1),1/2,1/2,1 \rangle &:& \phi_{\rho}(p) \;=\; 
[ |udu \rangle - |duu \rangle ]/\sqrt{2} ~, 
\nonumber\\
&:& \phi_{\lambda}(p) \;=\; 
[ 2|uud \rangle - |udu \rangle - |duu \rangle ]/\sqrt{6} ~, 
\nonumber\\
|(1,1),1,1,0 \rangle &:& \phi_{\rho}(\Sigma^+) \;=\; 
[ |suu \rangle - |usu \rangle ]/\sqrt{2} ~, 
\nonumber\\
&:& \phi_{\lambda}(\Sigma^+) \;=\; 
[ |suu \rangle + |usu \rangle - 2|uus \rangle ]/\sqrt{6} ~, 
\nonumber\\
|(1,1),0,0,0 \rangle &:& \phi_{\rho}(\Lambda) \;=\; 
[ 2|uds \rangle - 2|dus \rangle - |dsu \rangle 
+ |sdu \rangle - |sud \rangle + |usd \rangle ]/\sqrt{12} ~, 
\nonumber\\
&:& \phi_{\lambda}(\Lambda) \;=\; 
[- |dsu \rangle - |sdu \rangle + |sud \rangle + |usd \rangle ]/2 ~, 
\nonumber\\
|(1,1),1/2,1/2,-1 \rangle &:& \phi_{\rho}(\Xi^0) \;=\; 
[ |sus \rangle - |uss \rangle ]/\sqrt{2} ~, 
\nonumber\\
&:& \phi_{\lambda}(\Xi^0) \;=\; 
[ 2|ssu \rangle - |sus \rangle - |uss \rangle ]/\sqrt{6} ~.
\ea
(ii) The decuplet baryons $(p,q)=(3,0)$:
\ba
|(3,0),3/2,3/2,1 \rangle 
&:& \phi_S(\Delta^{++}) \;=\; |uuu \rangle ~, 
\nonumber\\
|(3,0),1,1,0 \rangle &:& \phi_S(\Sigma^{+}) \;=\; 
[ |suu \rangle + |usu \rangle + |uus \rangle ]/\sqrt{3} ~, 
\nonumber\\
|(3,0),1/2,1/2,-1 \rangle &:& \phi_S(\Xi^{0}) \;=\; 
[ |ssu \rangle + |sus \rangle + |uss \rangle ]/\sqrt{3} ~, 
\nonumber\\
|(3,0),0,0,-2 \rangle  
&:& \phi_S(\Omega^{-}) \;=\; |sss \rangle ~. 
\ea
(iii) The singlet baryons $(p,q)=(0,0)$:
\ba
|(0,0),0,0,0 \rangle &:& \phi_A(\Lambda) \;=\; 
[ |uds \rangle - |dus \rangle + |dsu \rangle 
- |sdu \rangle + |sud \rangle - |usd \rangle ]/\sqrt{6} ~. 
\ea
We only show the highest charge state $M_I=I$ with $Q=I+Y/2$. 
The other charge states are obtained by applying the lowering operator 
in isospin space.

\section{Baryon wave functions}
\setcounter{equation}{0}

The $S_3$ invariant space-spin-flavor ($\Psi=\psi \chi \phi$) 
baryon wave functions are given by 
\ba
^{2}8[56,L^P] &:& \psi_S ( \chi_{\rho} \phi_{\rho} 
+ \chi_{\lambda} \phi_{\lambda})/\sqrt{2} ~,
\nonumber\\
^{2}8[70,L^P] &:& [ \psi_{\rho} ( \chi_{\rho} \phi_{\lambda} 
+ \chi_{\lambda} \phi_{\rho}) + \psi_{\lambda} ( \chi_{\rho} \phi_{\rho} 
- \chi_{\lambda} \phi_{\lambda} ) ]/2 ~, 
\nonumber\\
^{4}8[70,L^P] &:& ( \psi_{\rho} \phi_{\rho} 
+ \psi_{\lambda} \phi_{\lambda} ) \chi_S/\sqrt{2} ~,
\nonumber\\
^{2}8[20,L^P] &:& \psi_A ( \chi_{\rho} \phi_{\lambda} 
- \chi_{\lambda} \phi_{\rho} )/\sqrt{2} ~,
\nonumber\\
^{4}10[56,L^P] &:& \psi_S \chi_S \phi_S ~,
\nonumber\\
^{2}10[70,L^P] &:& ( \psi_{\rho} \chi_{\rho} 
+ \psi_{\lambda} \chi_{\lambda} ) \phi_S/\sqrt{2} ~,
\nonumber\\
^{2}1[70,L^P] &:& ( \psi_{\rho} \chi_{\lambda} 
- \psi_{\lambda} \chi_{\rho} ) \phi_A/\sqrt{2} ~,
\nonumber\\
^{4}1[20,L^P] &:& \psi_A \chi_S \phi_A ~. 
\ea
The quark orbital angular momentum $L$ is coupled with the spin 
$S$ to the total angular momentum $J$ of the baryon.

\clearpage

\begin{table}
\centering
\caption[]{Classification of ground state baryons.}
\label{iy}
\vspace{15pt}
\begin{tabular}{lccr}
\hline
& & & \\
& Baryon & I & Y \\
& & & \\
\hline
& & & \\
$J^P=\frac{1}{2}^+$ octet: 
& $N$       & 1/2 &   1 \\
& $\Sigma$  & 1   &   0 \\
& $\Lambda$ & 0   &   0 \\
& $\Xi$     & 1/2 & --1 \\
$J^P=\frac{3}{2}^+$ decuplet: 
& $\Delta$        & 3/2 &   1 \\
& $\Sigma^{*}$ & 1   &   0 \\
& $\Xi^{*}$    & 1/2 & --1 \\
& $\Omega$        & 0   & --2 \\
$J^P=\frac{1}{2}^+$ singlet:   
& $\Lambda^{*}$ & 0   &  0 \\
& & & \\
\hline
\end{tabular}
\end{table}

\clearpage

\begin{table}
\centering
\caption[]{Values of the parameters in the mass 
formula of Eq.~(\protect\ref{massformula}) in GeV$^2$.}
\label{par}
\vspace{15pt}
\begin{tabular}{crr}
\hline
& & \\
Parameter & Present & Ref.~\protect\cite{BIL} \\
& & \\
\hline
& & \\
$M_0^2$    &   0.882 &   0.882 \\
$\kappa_1$ &   1.204 &   1.192 \\
$\kappa_2$ &   1.460 &   1.535 \\
$\alpha$   &   1.068 &   1.064 \\
$a$        & --0.041 & --0.042 \\
$b$        &   0.017 &   0.030 \\
$c$        &   0.130 &   0.124 \\
$d$        & --0.449 & \\
$e$        &   0.016 & \\
$f$        &   0.042 & \\
& & \\
$\delta$(MeV) & 33 & 39 \\
& & \\
\hline
\end{tabular}
\end{table}

\clearpage

\begin{table}
\centering
\caption[]{Mass spectrum of nonstrange baryon resonances 
in the oblate top model. The masses are given in MeV. 
The experimental values are taken from \protect\cite{PDG}.}
\label{nucdel} 
\vspace{15pt} 
\begin{tabular}{llcccr}
\hline
& & & & & \\
Baryon $L_{2I,2J}$ & Status & Mass 
& State & ($v_1,v_2$) & $M_{\mbox{calc}}$ \\
& & & & & \\
\hline
& & & & & \\
$N( 939)P_{11}$   & **** & 939       
& $^{2}8_{ 1/2}[56,0^+]$ & (0,0) &  939 \\
$N(1440)P_{11}$   & **** & 1430-1470 
& $^{2}8_{ 1/2}[56,0^+]$ & (1,0) & 1444 \\
$N(1520)D_{13}$   & **** & 1515-1530 
& $^{2}8_{ 3/2}[70,1^-]$ & (0,0) & 1563 \\
$N(1535)S_{11}$   & **** & 1520-1555 
& $^{2}8_{ 1/2}[70,1^-]$ & (0,0) & 1563 \\
$N(1650)S_{11}$   & **** & 1640-1680 
& $^{4}8_{ 1/2}[70,1^-]$ & (0,0) & 1683 \\
$N(1675)D_{15}$   & **** & 1670-1685 
& $^{4}8_{ 5/2}[70,1^-]$ & (0,0) & 1683 \\
$N(1680)F_{15}$   & **** & 1675-1690 
& $^{2}8_{ 5/2}[56,2^+]$ & (0,0) & 1737 \\
$N(1700)D_{13}$   &  *** & 1650-1750 
& $^{4}8_{ 3/2}[70,1^-]$ & (0,0) & 1683 \\
$N(1710)P_{11}$   &  *** & 1680-1740 
& $^{2}8_{ 1/2}[70,0^+]$ & (0,1) & 1683 \\
$N(1720)P_{13}$   & **** & 1650-1750 
& $^{2}8_{ 3/2}[56,2^+]$ & (0,0) & 1737 \\
$N(2190)G_{17}$   & **** & 2100-2200 
& $^{2}8_{ 7/2}[70,3^-]$ & (0,0) & 2140 \\
$N(2220)H_{19}$   & **** & 2180-2310 
& $^{2}8_{ 9/2}[56,4^+]$ & (0,0) & 2271 \\
$N(2250)G_{19}$   & **** & 2170-2310 
& $^{4}8_{ 9/2}[70,3^-]$ & (0,0) & 2229 \\
$N(2600)I_{1,11}$ &  *** & 2550-2750 
& $^{2}8_{11/2}[70,5^-]$ & (0,0) & 2591 \\
& & & & & \\
$\Delta(1232)P_{33}$ & **** & 1230-1234  
& $^{4}10_{3/2}[56,0^+]$ & (0,0) & 1246 \\
$\Delta(1600)P_{33}$ &  *** & 1550-1700 
& $^{4}10_{3/2}[56,0^+]$ & (1,0) & 1660 \\
$\Delta(1620)S_{31}$ & **** & 1615-1675 
& $^{2}10_{1/2}[70,1^-]$ & (0,0) & 1649 \\
$\Delta(1700)D_{33}$ & **** & 1670-1770 
& $^{2}10_{3/2}[70,1^-]$ & (0,0) & 1649 \\
$\Delta(1905)F_{35}$ & **** & 1870-1920 
& $^{4}10_{5/2}[56,2^+]$ & (0,0) & 1921 \\ 
$\Delta(1910)P_{31}$ & **** & 1870-1920 
& $^{4}10_{1/2}[56,2^+]$ & (0,0) & 1921 \\ 
$\Delta(1920)P_{33}$ &  *** & 1900-1970 
& $^{4}10_{3/2}[56,2^+]$ & (0,0) & 1921 \\ 
$\Delta(1930)D_{35}$ &  *** & 1920-1970 
& $^{2}10_{5/2}[70,2^-]$ & (0,0) & 1946 \\ 
$\Delta(1950)F_{37}$ & **** & 1940-1960 
& $^{4}10_{7/2}[56,2^+]$ & (0,0) & 1921 \\ 
$\Delta(2420)H_{3,11}$ & **** & 2300-2500 
& $^{4}10_{11/2}[56,4^+]$ & (0,0) & 2414 \\
& & & & & \\
\hline
\end{tabular}
\end{table}

\clearpage

\begin{table}
\centering
\caption[]{As Table~\protect\ref{nucdel}, but for strange 
baryon resonances. Note, that $\Xi$ resonances are denoted by 
$L_{2I,2J}$.}  
\label{strange} 
\vspace{15pt} 
\begin{tabular}{llcccr}
\hline
& & & & & \\
Baryon $L_{I,2J}$ & Status & Mass 
& State & ($v_1,v_2$) & $M_{\mbox{calc}}$ \\
& & & & & \\
\hline
& & & & & \\
$\Sigma(1193)P_{11}$   & **** &      1193  
& $^{2}8_{ 1/2}[56,0^+]$ & (0,0) & 1170 \\
$\Sigma(1660)P_{11}$   &  *** & 1630-1690 
& $^{2}8_{ 1/2}[56,0^+]$ & (1,0) & 1604 \\
$\Sigma(1670)D_{13}$   & **** & 1665-1685 
& $^{2}8_{ 3/2}[70,1^-]$ & (0,0) & 1711 \\
$\Sigma(1750)S_{11}$   &  *** & 1730-1800 
& $^{2}8_{ 1/2}[70,1^-]$ & (0,0) & 1711 \\
$\Sigma(1775)D_{15}$   & **** & 1770-1780 
& $^{4}8_{ 5/2}[70,1^-]$ & (0,0) & 1822 \\
$\Sigma(1915)F_{15}$   & **** & 1900-1935 
& $^{2}8_{ 5/2}[56,2^+]$ & (0,0) & 1872 \\
$\Sigma(1940)D_{13}$   &  *** & 1900-1950 
& $^{2}8_{ 3/2}[56,1^-]$ & (0,1) & 1974 \\
$\Sigma^*(1385)P_{13}$   & **** & 1383-1385 
& $^{4}10_{ 3/2}[56,0^+]$ & (0,0) & 1382 \\
$\Sigma^*(2030)F_{17}$   & **** & 2025-2040 
& $^{4}10_{ 7/2}[56,2^+]$ & (0,0) & 2012 \\
& & & & & \\
$\Lambda(1116)P_{01}$   & **** & 1116      
& $^{2}8_{ 1/2}[56,0^+]$ & (0,0) & 1133 \\
$\Lambda(1600)P_{01}$   &  *** & 1560-1700 
& $^{2}8_{ 1/2}[56,0^+]$ & (1,0) & 1577 \\
$\Lambda(1670)S_{01}$   & **** & 1660-1680 
& $^{2}8_{ 1/2}[70,1^-]$ & (0,0) & 1686 \\
$\Lambda(1690)D_{03}$   & **** & 1685-1690 
& $^{2}8_{ 3/2}[70,1^-]$ & (0,0) & 1686 \\
$\Lambda(1800)S_{01}$   &  *** & 1720-1850 
& $^{4}8_{ 1/2}[70,1^-]$ & (0,0) & 1799 \\
$\Lambda(1810)P_{01}$   &  *** & 1750-1850 
& $^{2}8_{ 1/2}[70,0^+]$ & (0,1) & 1799 \\
$\Lambda(1820)F_{05}$   & **** & 1815-1825 
& $^{2}8_{ 5/2}[56,2^+]$ & (0,0) & 1849 \\
$\Lambda(1830)D_{05}$   & **** & 1810-1830 
& $^{4}8_{ 5/2}[70,1^-]$ & (0,0) & 1799 \\
$\Lambda(1890)P_{03}$   & **** & 1850-1910 
& $^{2}8_{ 3/2}[56,2^+]$ & (0,0) & 1849 \\
$\Lambda(2110)F_{05}$   & **** & 2090-2140 
& $^{4}8_{ 5/2}[70,2^+]$ & (0,0) & 2074 \\
$\Lambda(2350)H_{09}$   &  *** & 2340-2370 
& $^{2}8_{ 9/2}[56,4^+]$ & (0,0) & 2357 \\
$\Lambda^*(1405)S_{01}$   & **** & 1402-1410 
& $^{2}1_{ 1/2}[70,1^-]$ & (0,0) & 1641 \\
$\Lambda^*(1520)D_{03}$   & **** & 1518-1520 
& $^{2}1_{ 3/2}[70,1^-]$ & (0,0) & 1641 \\
$\Lambda^*(2100)G_{07}$   & **** & 2090-2110 
& $^{2}1_{ 7/2}[70,3^-]$ & (0,0) & 2197 \\
& & & & & \\
$\Xi(1318)P_{11}$    & **** & 1314-1316 
& $^{2} 8_{1/2}[56,0^+]$ & (0,0) & 1334 \\
$\Xi(1820)D_{13}$    &  *** & 1818-1828 
& $^{2} 8_{3/2}[70,1^-]$ & (0,0) & 1828 \\
$\Xi^*(1530)P_{13}$    & **** & 1531-1532 
& $^{4}10_{3/2}[56,0^+]$ & (0,0) & 1524 \\
& & & & & \\
$\Omega(1672)P_{03}$ & **** & 1672-1673 
& $^{4}10_{3/2}[56,0^+]$ & (0,0) & 1670 \\
& & & & & \\
\hline
\end{tabular}
\end{table}

\clearpage

\begin{table}
\centering
\caption[]{Masses of low-lying octet baryons in MeV. Assignments 
of three and four star resonances are labeled by $^{\dagger}$, 
and tentative assignments of one and two star resonances by 
$^{\ddagger}$.} 
\label{miss8} 
\vspace{15pt} 
\begin{tabular}{ccllll}
\hline
& & & & & \\
State & ($v_1,v_2$) & $\;$ $N$ & $\;$ $\Sigma$ & $\;$ $\Lambda$ 
& $\;$ $\Xi$ \\
& & & & & \\
\hline
& & & & & \\
$^{2}8_J[56,0^+]$ & (0,0) & $\,$ 939 $^{\dagger}$ & 1170 $^{\dagger}$ 
& 1133 $^{\dagger}$ & 1334 $^{\dagger}$ \\
$^{2}8_J[70,1^-]$ & (0,0) & 1563 $^{\dagger}$ & 1711 $^{\dagger}$ 
& 1686 $^{\dagger}$ & 1828 $^{\dagger}$ \\ 
$^{4}8_J[70,1^-]$ & (0,0) & 1683 $^{\dagger}$ & 1822 $^{\dagger}$ 
& 1799 $^{\dagger}$ & 1932 \\
$^{2}8_J[20,1^+]$ & (0,0) & 1713 & 1849 & 1826 & 1957 \\
$^{2}8_J[56,2^+]$ & (0,0) & 1737 $^{\dagger}$ & 1872 $^{\dagger}$ 
& 1849 $^{\dagger}$ & 1979 \\
$^{2}8_J[70,2^+]$ & (0,0) & 1874 $^{\ddagger}$ & 1999 
& 1978 & 2100 \\
$^{2}8_J[70,2^-]$ & (0,0) & 1874 & 1999 & 1978 & 2100 \\
$^{4}8_J[70,2^+]$ & (0,0) & 1975 $^{\ddagger}$ & 2095 
& 2074 $^{\dagger}$ & 2191 \\
$^{4}8_J[70,2^-]$ & (0,0) & 1975 & 2095 & 2074 & 2191 \\
& & & & & \\
$^{2}8_J[56,0^+]$ & (1,0) & 1444 $^{\dagger}$ & 1604 $^{\dagger}$ 
& 1577 $^{\dagger}$ & 1727 \\
$^{2}8_J[70,1^-]$ & (1,0) & 1909 & 2033 & 2012 & 2132 \\
$^{4}8_J[70,1^-]$ & (1,0) & 2009 & 2127 & 2107 & 2222 \\
$^{2}8_J[20,1^+]$ & (1,0) & 2034 & 2150 & 2130 & 2244 \\
& & & & & \\
$^{2}8_J[70,0^+]$ & (0,1) & 1683 $^{\dagger}$ & 1822 $^{\ddagger}$ 
& 1799 $^{\dagger}$ & 1932 \\
$^{4}8_J[70,0^+]$ & (0,1) & 1796 & 1926 & 1904 & 2030 \\
$^{2}8_J[56,1^-]$ & (0,1) & 1847 & 1974 $^{\dagger}$ 
& 1952 & 2076 \\
$^{2}8_J[70,1^-]$ & (0,1) & 1975 & 2095 & 2074 & 2191 \\
$^{2}8_J[70,1^+]$ & (0,1) & 1975 & 2095 & 2074 & 2191 \\
$^{4}8_J[70,1^-]$ & (0,1) & 2072 & 2186 & 2167 & 2278 \\
$^{4}8_J[70,1^+]$ & (0,1) & 2072 & 2186 & 2167 & 2278 \\
$^{2}8_J[20,1^-]$ & (0,1) & 2096 & 2209 & 2190 & 2300 \\
& & & & & \\
\hline
\end{tabular}
\end{table}

\clearpage

\begin{table}
\centering
\caption[]{As Table~\protect\ref{miss8}, but for decuplet baryons.} 
\label{miss10} 
\vspace{15pt} 
\begin{tabular}{ccllll}
\hline
& & & & & \\
State & ($v_1,v_2$) & $\;$ $\Delta$ & $\;$ $\Sigma^{*}$ & 
$\;$ $\Xi^{*}$ & $\;$ $\Omega$ \\
& & & & & \\
\hline
& & & & & \\
$^{4}10_J[56,0^+]$ & (0,0) & 1246 $^{\dagger}$ 
& 1382 $^{\dagger}$ & 1524 $^{\dagger}$ & 1670 $^{\dagger}$ \\
$^{2}10_J[70,1^-]$ & (0,0) & 1649 $^{\dagger}$ 
& 1755 & 1869 & 1989 \\
$^{4}10_J[56,2^+]$ & (0,0) & 1921 $^{\dagger}$ 
& 2012 $^{\dagger}$ & 2112 & 2219 \\
$^{2}10_J[70,2^+]$ & (0,0) & 1946 $^{\ddagger}$ 
& 2037 & 2135 & 2242 \\
$^{2}10_J[70,2^-]$ & (0,0) & 1946 $^{\dagger}$ 
& 2037 & 2135 & 2242 \\
& & & & & \\
$^{4}10_J[56,0^+]$ & (1,0) & 1660 $^{\dagger}$ 
& 1765 & 1878 & 1998 \\
$^{2}10_J[70,1^-]$ & (1,0) & 1981 $^{\ddagger}$ 
& 2070 & 2167 & 2272 \\
& & & & & \\
$^{2}10_J[70,0^+]$ & (0,1) & 1764 $^{\ddagger}$ 
& 1863 & 1970 & 2085 \\
$^{4}10_J[56,1^-]$ & (0,1) & 2020 & 2107 & 2203 & 2306 \\
$^{2}10_J[70,1^-]$ & (0,1) & 2044 & 2131 & 2225 & 2327 \\
$^{2}10_J[70,1^+]$ & (0,1) & 2044 & 2131 & 2225 & 2327 \\
& & & & & \\
\hline
\end{tabular}
\end{table}

\clearpage

\begin{table}
\centering
\caption[]{As Table~\protect\ref{miss8}, but for singlet baryons.} 
\label{miss1} 
\vspace{15pt} 
\begin{tabular}{ccl}
\hline
& & \\
State & ($v_1,v_2$) & $\;$ $\Lambda^{*}$ \\
& & \\
\hline
& & \\
$^{2}1_J[70,1^-]$ & (0,0) & 1641 $^{\dagger}$ \\
$^{4}1_J[20,1^+]$ & (0,0) & 1891 \\
$^{2}1_J[70,2^+]$ & (0,0) & 1939 \\
$^{2}1_J[70,2^-]$ & (0,0) & 1939 \\
& & \\
$^{2}1_J[70,1^-]$ & (1,0) & 1974 \\ 
$^{4}1_J[20,1^+]$ & (1,0) & 2186 \\
& & \\
$^{2}1_J[70,0^+]$ & (0,1) & 1756 \\
$^{2}1_J[70,1^-]$ & (0,1) & 2038 \\
$^{2}1_J[70,1^+]$ & (0,1) & 2038 \\
$^{4}1_J[20,1^-]$ & (0,1) & 2244 \\
& & \\
\hline
\end{tabular}
\end{table}

\clearpage

\begin{table}
\centering
\caption[]{Collective form factors in the large $N$ limit. 
The states are labeled by $[dim,L^P]_{(v_1,v_2)}$, where $dim$
denotes the dimension of the $SU_{\rm sf}(6)$ representation.
The final state is $[56,0^+]_{(0,0)}$. The form factors for 
vibrational excitations are proportional to the coefficients 
$\chi_1$ and $\chi_2$ \cite{strong}. $H(x)=\arctan x - x/(1+x^2)$.}
\label{collff} 
\vspace{15pt}
\begin{tabular}{cc}
\hline
& \\
Initial state & ${\cal F}(k)$ \\
& \\
\hline
& \\
$[56,0^+]_{(0,0)}$
& $\frac{1}{(1+k^2a^2)^2}$ \\
& \\
$[20,1^+]_{(0,0)}$ & 0 \\
& \\
$[70,1^-]_{(0,0)}$ 
& $i \, \sqrt{3} \, \frac{ka}{(1+k^2a^2)^2}$ \\
& \\
$[56,2^+]_{(0,0)}$
& $\frac{1}{2}\sqrt{5}\left[ \frac{-1}{(1+k^2a^2)^2} 
+ \frac{3}{2k^3a^3} H(ka) \right]$ \\
& \\
$[70,2^-]_{(0,0)}$ & 0 \\
& \\
$[70,2^+]_{(0,0)}$
& $-\frac{1}{2}\sqrt{15}\left[ \frac{-1}{(1+k^2a^2)^2} 
+ \frac{3}{2k^3a^3} H(ka) \right]$ \\
& \\
$[56,0^+]_{(1,0)}$
& $-\chi_1 \frac{2k^2a^2}{(1+k^2a^2)^3}$ \\
& \\
$[20,1^+]_{(1,0)}$ & 0 \\
& \\
$[70,1^-]_{(1,0)}$ 
& $i \chi_1 \sqrt{3} \, \frac{ka(1-3k^2a^2)}{2(1+k^2a^2)^3}$ \\
& \\
$[70,0^+]_{(0,1)}$
& $ \chi_2 \frac{2k^2a^2}{(1+k^2a^2)^3}$ \\
& \\
$[70,1^+]_{(0,1)}$ & 0 \\
& \\
$[56,1^-]_{(0,1)}$ 
& $-i\chi_2 \sqrt{6} \frac{k^3a^3}{(1+k^2a^2)^3}$ \\
& \\
$[20,1^-]_{(0,1)}$ & 0 \\
& \\
$[70,1^-]_{(0,1)}$ & $-i\chi_2 \sqrt{\frac{3}{2}} 
\frac{ka(1-k^2a^2)}{(1+k^2a^2)^3}$ \\
& \\
\hline
\end{tabular}
\end{table}

\clearpage

\begin{table}
\centering
\caption[]{Coefficients $\alpha_{m,\gamma}$ ($m=0,\pm$) of 
Eq.~(\protect\ref{alfa}) for 
strong decay of baryons $B \rightarrow B_8 + M_8$ where 
$B_8$ is an octet ground state baryon with $^{2}8[56]$ 
and $M_8$ an octet meson.}
\label{B8M8} 
\vspace{15pt}
\begin{tabular}{c|cccccc}
\hline
& & & & & & \\
& \multicolumn{6}{c}{Helicity $\nu=1/2$} \\
B & \multicolumn{2}{c}{$\alpha_{-,\gamma}$} 
  & \multicolumn{2}{c}{$\alpha_{0,\gamma}$} 
  & \multicolumn{2}{c}{$\alpha_{+,\gamma}$} \\
& & & & & & \\
\hline
& & & & & & \\
$^{2}8[56]$ & 0 & 0 & $\frac{\sqrt{5}}{3\sqrt{3}}$ 
& $\frac{2}{3\sqrt{3}}$ & $\frac{2\sqrt{5}}{3\sqrt{3}}$ 
& $\frac{4}{3\sqrt{3}}$ \\
& & & & & & \\
$^{2}8[70]$ & 0 & 0 & $\frac{\sqrt{5}}{6\sqrt{6}}$ 
& $\frac{5}{6\sqrt{6}}$ & $\frac{\sqrt{5}}{3\sqrt{6}}$ 
& $\frac{5}{3\sqrt{6}}$ \\
& & & & & & \\
$^{4}8[70]$ & $-\frac{\sqrt{5}}{3\sqrt{2}}$ 
& $\frac{1}{3\sqrt{2}}$ & $\frac{\sqrt{5}}{3\sqrt{6}}$ 
& $-\frac{1}{3\sqrt{6}}$ & $\frac{\sqrt{5}}{3\sqrt{6}}$ 
& $-\frac{1}{3\sqrt{6}}$ \\
& & & & & & \\
$^{2}8[20]$ & 0 & 0 & 0 & 0 & 0 & 0 \\
& & & & & & \\
$^{4}10[56]$ & \multicolumn{2}{c}{$\frac{2\sqrt{2}}{3}$} 
& \multicolumn{2}{c}{$-\frac{2\sqrt{2}}{3\sqrt{3}}$} 
& \multicolumn{2}{c}{$-\frac{2\sqrt{2}}{3\sqrt{3}}$} \\
& & & & & & \\
$^{2}10[70]$ & \multicolumn{2}{c}{0} 
& \multicolumn{2}{c}{$-\frac{1}{3\sqrt{6}}$} 
& \multicolumn{2}{c}{$-\frac{\sqrt{2}}{3\sqrt{3}}$} \\
& & & & & & \\
$^{2}1[70]$ & \multicolumn{2}{c}{0} 
& \multicolumn{2}{c}{$-\frac{1}{\sqrt{3}}$} 
& \multicolumn{2}{c}{$-\frac{2}{\sqrt{3}}$} \\
& & & & & & \\
$^{4}1[20]$ & \multicolumn{2}{c}{0} & \multicolumn{2}{c}{0} 
& \multicolumn{2}{c}{0} \\
& & & & & & \\
\hline
\end{tabular}
\end{table}

\clearpage

\begin{table}
\centering
\caption[]{Coefficients $\alpha_{m,\gamma}$ ($m=0,\pm$) of 
Eq.~(\protect\ref{alfa}) for 
strong decay of baryons $B \rightarrow B_{10} + M_8$ where 
$B_{10}$ is an decuplet ground state baryon with $^{4}10[56]$ 
and $M_8$ an octet meson.}
\label{B10M8} 
\vspace{15pt}
\begin{tabular}{c|ccc|ccc}
\hline
& & & & & & \\
& \multicolumn{3}{c|} {Helicity $\nu=1/2$} 
& \multicolumn{3}{c} {Helicity $\nu=3/2$} \\
B & $\alpha_{-,\gamma}$ & $\alpha_{0,\gamma}$ & $\alpha_{+,\gamma}$ 
& $\alpha_{-,\gamma}$ & $\alpha_{0,\gamma}$ & $\alpha_{+,\gamma}$ \\
& & & & & & \\
\hline
& & & & & & \\
$^{2}8[56]$ & 0 & $\frac{\sqrt{10}}{3\sqrt{3}}$ 
& $-\frac{\sqrt{10}}{3\sqrt{3}}$ 
& 0 & 0 & $-\frac{\sqrt{10}}{3}$ \\
& & & & & & \\
$^{2}8[70]$ & 0 & $-\frac{\sqrt{5}}{3\sqrt{3}}$ 
& $\frac{\sqrt{5}}{3\sqrt{3}}$ 
& 0 & 0 & $\frac{\sqrt{5}}{3}$ \\
& & & & & & \\
$^{4}8[70]$ & $-\frac{\sqrt{5}}{3}$ & $-\frac{\sqrt{5}}{6\sqrt{3}}$ 
& $-\frac{2\sqrt{5}}{3\sqrt{3}}$ 
& 0 & $-\frac{\sqrt{5}}{2\sqrt{3}}$ & $-\frac{\sqrt{5}}{3}$ \\
& & & & & & \\
$^{2}8[20]$ & 0 & 0 & 0 & 0 & 0 & 0 \\
& & & & & & \\
$^{4}10[56]$ & $\frac{2\sqrt{2}}{3}$ & $\frac{\sqrt{2}}{3\sqrt{3}}$ 
& $\frac{4\sqrt{2}}{3\sqrt{3}}$ 
& 0 & $\frac{\sqrt{2}}{\sqrt{3}}$ & $\frac{2\sqrt{2}}{3}$ \\
& & & & & & \\
$^{2}10[70]$ & 0 & $-\frac{2\sqrt{2}}{3\sqrt{3}}$ 
& $\frac{2\sqrt{2}}{3\sqrt{3}}$ 
& 0 & 0 & $\frac{2\sqrt{2}}{3}$ \\
& & & & & & \\
\hline
\end{tabular}
\end{table}

\clearpage

\begin{table}
\centering
\caption[]{Coefficients $\alpha_{m,\gamma}$ ($m=0,\pm$) of 
Eq.~(\protect\ref{alfa}) for 
strong decay of baryons $B \rightarrow B_8 + M_1$ where 
$B_8$ is an octet ground state baryon with $^{2}8[56]$ 
and $M_1$ a singlet meson.}
\label{B8M1} 
\vspace{15pt}
\begin{tabular}{c|ccc}
\hline
& & & \\
& \multicolumn{3}{c}{Helicity $\nu=1/2$} \\
B & $\alpha_{-,\gamma}$ & $\alpha_{0,\gamma}$ & $\alpha_{+,\gamma}$ \\
& & & \\
\hline
& & & \\
$^{2}8[56]$ & 0 & $\frac{1}{3\sqrt{6}}$ & $\frac{\sqrt{2}}{3\sqrt{3}}$ \\
& & & \\
$^{2}8[70]$ & 0 & $\frac{1}{3\sqrt{3}}$ & $\frac{2}{3\sqrt{3}}$ \\
& & & \\
$^{4}8[70]$ & $\frac{1}{3}$ & $-\frac{1}{3\sqrt{3}}$ 
& $-\frac{1}{3\sqrt{3}}$ \\
& & & \\
$^{2}8[20]$ & 0 & 0 & 0 \\
& & & \\
\hline
\end{tabular}
\end{table}

\clearpage

\begin{table}
\centering
\caption[]{Coefficients $\alpha_{m,\gamma}$ ($m=0,\pm$) of 
Eq.~(\protect\ref{alfa}) for 
strong decay of baryons $B \rightarrow B_{10} + M_{1}$ where 
$B_{10}$ is an decuplet ground state baryon with $^{4}10[56]$ 
and $M_{1}$ a singlet meson.}
\label{B10M1} 
\vspace{15pt}
\begin{tabular}{c|ccc|ccc}
\hline
& & & & & & \\
& \multicolumn{3}{c|} {Helicity $\nu=1/2$} 
& \multicolumn{3}{c} {Helicity $\nu=3/2$} \\
B & $\alpha_{-,\gamma}$ & $\alpha_{0,\gamma}$ & $\alpha_{+,\gamma}$ 
& $\alpha_{-,\gamma}$ & $\alpha_{0,\gamma}$ & $\alpha_{+,\gamma}$ \\
& & & & & & \\
\hline
& & & & & & \\
$^{4}10[56]$ & $\frac{\sqrt{2}}{3}$ & $\frac{1}{3\sqrt{6}}$ 
& $\frac{2\sqrt{2}}{3\sqrt{3}}$ 
& 0 & $\frac{1}{\sqrt{6}}$ & $\frac{\sqrt{2}}{3}$ \\
& & & & & & \\
$^{2}10[70]$ & 0 & $-\frac{\sqrt{2}}{3\sqrt{3}}$ 
& $\frac{\sqrt{2}}{3\sqrt{3}}$ 
& 0 & 0 & $\frac{\sqrt{2}}{3}$ \\
& & & & & & \\
\hline
\end{tabular}
\end{table}

\clearpage

\begin{table}
\centering
\caption[]{Strong decay widths of three and four star nucleon 
resonances in MeV. 
The experimental values are taken from \protect\cite{PDG}. 
Decay channels labeled by -- are below threshold.}
\label{nuc} 
\vspace{15pt}
\begin{tabular}{l|cccc|cc}
\hline
& & & & & & \\
Baryon & $N \pi$ & $N \eta$ & $\Sigma K$ & $\Lambda K$ 
& $\Delta \pi$ & $\Sigma^* K$ \\
& & & & & & \\
\hline
& & & & & & \\
$N(1440)P_{11}$ & 108 & --  & -- & -- & 0 & -- \\
& $227 \pm 67$ & & & & $87 \pm 30$ & \\
$N(1520)D_{13}$ & 115 & 1 & -- & --  &  12 & -- \\
& $67 \pm 9$ & & & & $24 \pm 7$ & \\
$N(1535)S_{11}$ &  85 & 0 & -- & --  &  23 & -- \\
& $79 \pm 38$ & $74 \pm 39$ & & & $< 1 \pm 1$ & \\
$N(1650)S_{11}$ &  35 &   8 & -- & 0 &  24 & -- \\
& $121 \pm 34$ & $11 \pm 6$ & & $12 \pm 7$ & $7 \pm 5$ & \\
$N(1675)D_{15}$ &  31 & 17  & -- & 0   & 123 & -- \\
& $72 \pm 12$ & & & $< 1 \pm 1$ & $88 \pm 14$ & \\
$N(1680)F_{15}$ &  41 & 0 & -- & 0 &   5 & -- \\
& $84 \pm 9$ & & & & $13 \pm 7$ & \\
$N(1700)D_{13}$ &   5 & 4   & 0 & 0 & 225 & -- \\
& $10 \pm 7$ & & & $< 2 \pm 2$ & & \\
$N(1710)P_{11}$ & 85 &  8 & 0 & 1 & 34 & -- \\
& $23 \pm 17$ & & & $23 \pm 21$ & $41 \pm 33$ & \\
$N(1720)P_{13}$ & 31 & 0 & 0 & 0 & 10 & -- \\
& $23 \pm 11$ & & & $12 \pm 11$ & & \\
$N(2190)G_{17}$ & 34 &  11 & 1 & 7 & 25 & 1 \\
& $68 \pm 27$ & & & & & \\
$N(2220)H_{19}$ & 15 & 1 & 0 & 2 &  5 & 0 \\
& $65 \pm 28$ & & & & & \\
$N(2250)G_{19}$ &  7 &   9 &   9 & 0 & 40 & 2 \\
& $38 \pm 21$ & & & & & \\
$N(2600)I_{1,11}$ & 9 &   3 & 0 & 3 &  7 & 1 \\
& $49 \pm 20$ & & & & & \\
& & & & & & \\
\hline
\end{tabular}
\end{table}

\clearpage

\begin{table}
\centering
\caption[]{As Table~\protect\ref{nuc}, but for $\Delta$ resonances.}
\label{del} 
\vspace{15pt}
\begin{tabular}{l|cc|ccc}
\hline
& & & & & \\
Baryon & $N \pi$ & $\Sigma K$ 
& $\Delta \pi$ & $\Delta \eta$ & $\Sigma^* K$ \\
& & & & & \\
\hline
& & & & & \\
$\Delta(1232)P_{33}$ & 116 & -- & -- & -- & -- \\
& $119 \pm 5$ & & & & \\
$\Delta(1600)P_{33}$ & 108 & -- & 25 & -- & -- \\
& $61 \pm 32$ & & $193 \pm 76$ & & \\
$\Delta(1620)S_{31}$ &  16 & -- & 89 & -- & -- \\
& $38 \pm 11$ & & $68 \pm 26$ & & \\
$\Delta(1700)D_{33}$ &  27 & 0 & 144 & -- & -- \\
& $45 \pm 21$ & & $135 \pm 64$ & & \\
$\Delta(1905)F_{35}$ &   9 & 1 &  45 & 1 & 0 \\
& $36 \pm 20$ & & $< 45 \pm 45$ & & \\
$\Delta(1910)P_{31}$ &  42 &   2 &   4 & 0 & 0 \\
& $52 \pm 19$ & & & & \\
$\Delta(1920)P_{33}$ & 22 &   1 &  29 & 1 & 0 \\
& $28 \pm 19$ & & & & \\
$\Delta(1930)D_{35}$ &   0 &   0 &   0 &   0 &   0 \\
& $53 \pm 23$ & & & & \\
$\Delta(1950)F_{37}$ & 45 &   6 &  36 &   2 & 0 \\
& $120 \pm 14$ & & $80 \pm 18$ & & \\
$\Delta(2420)H_{3,11}$ & 12 &  4 &  11 &   2 &   1 \\
& $40 \pm 22$ & & & & \\
& & & & & \\
\hline
\end{tabular}
\end{table}

\clearpage

\begin{table}
\centering
\caption[]{As Table~\protect\ref{nuc}, but for $\Sigma$ resonances.}
\label{sig} 
\vspace{15pt}
\begin{tabular}{l|ccccc|cccc}
\hline
& & & & & & & & & \\
Baryon & $N \overline{K}$ & $\Sigma \pi$ & $\Lambda \pi$ 
& $\Sigma \eta$ & $\Xi K$ & $\Delta \overline{K}$ & $\Sigma^* \pi$ 
& $\Sigma^* \eta$ & $\Xi^* K$ \\
& & & & & & & & & \\
\hline
& & & & & & & & & \\
$\Sigma(1660)P_{11}$ & 2 & 40 & 29 & -- & -- 
& -- & 1 & -- & -- \\
& $24 \pm 20$ & seen & seen & & & & & & \\
$\Sigma(1670)D_{13}$ & 3 & 77 & 7 & -- & -- 
& -- &   2 & -- & -- \\
& $6 \pm 3$ & $27 \pm 13$ & $6 \pm 4$ & & & & & & \\
$\Sigma(1750)S_{11}$ & 3 & 85 & 7 & 0 & -- 
& 1 & 10 & -- & -- \\
& $28 \pm 21$ & $< 4 \pm 4$ & seen & $39 \pm 28$ & & & & & \\
$\Sigma(1775)D_{15}$ & 58 & 14 & 27 & 0 & -- 
&  4 & 14 & -- & -- \\
& $48 \pm 7$ & $4 \pm 2$ & $20 \pm 4$ & & & & $12 \pm 3$ & & \\
$\Sigma(1915)F_{15}$ &  1 & 20 & 10 & 1 & 1 
& 3 & 2 & -- & -- \\
& $12 \pm 7$ & seen & seen & & & & $< 3 \pm 3$ & & \\
$\Sigma(1940)D_{13}$ &  2 & 30 & 16 & 1 & 0  
& 4 & 3 & 0 & -- \\
& $< 23 \pm 23$ & seen & seen & & & seen & seen & & \\
$\Sigma^*(1385)P_{13}$ & -- & 9 & 49 & -- & -- 
& -- & -- & -- & -- \\
& & $4 \pm 1$ & $32 \pm 4$ & & & & & & \\
$\Sigma^*(2030)F_{17}$ & 12 & 11 & 19 & 5 & 1 
& 10 & 14 & 0 & 0 \\
& $35 \pm 7$ & $13 \pm 5$ & $35 \pm 7$ & & $< 2 \pm 2$ 
& $26 \pm 10$ & $18 \pm 9$ & & \\
& & & & & & & & & \\
\hline
\end{tabular}
\end{table}

\clearpage

\begin{table}
\centering
\caption[]{As Table~\protect\ref{nuc}, but for $\Lambda$ resonances.}
\label{lam} 
\vspace{15pt}
\begin{tabular}{l|cccc|cc}
\hline
& & & & & & \\
Baryon & $N \overline{K}$ & $\Sigma \pi$ & $\Lambda \eta$ 
& $\Xi K$ & $\Sigma^* \pi$ & $\Xi^* K$ \\
& & & & & & \\
\hline
& & & & & & \\
$\Lambda(1600)P_{01}$ & 25 & 21 & -- & -- & 0 & -- \\
& $34 \pm 25$ & $53 \pm 51$ & & & & \\
$\Lambda(1670)S_{01}$ & 44 & 9 & 0 & -- & 14 & -- \\
& $8 \pm 3$ & $15 \pm 9$ & $9 \pm 5$ & & & \\
$\Lambda(1690)D_{03}$ & 100 & 16 & 0 & -- & 14 & -- \\
& $15 \pm 4$ & $18 \pm 7$ & & & & \\
$\Lambda(1800)S_{01}$ &   0 & 80 &   5 & -- & 19 & -- \\
& $98 \pm 40$ & seen & & & seen & \\
$\Lambda(1810)P_{01}$ & 62 & 9 & 0 & -- & 15 & -- \\
& $53 \pm 42$ & $38 \pm 34$ & & & seen & \\
$\Lambda(1820)F_{05}$ & 23 & 13 & 0 & 0 & 3 & -- \\
& $48 \pm 7$ & $9 \pm 3$ & & & $6 \pm 2$ & \\
$\Lambda(1830)D_{05}$ &  0 & 77 &  16 & 0 & 101 & -- \\
& $6 \pm 3$ & $47 \pm 22$ & & & $> 13 \pm 4$ & \\
$\Lambda(1890)P_{03}$ & 19 & 12 & 0 & 0 &  10 & -- \\
& $36 \pm 22$ & $8 \pm 6$ & & & seen & \\
$\Lambda(2110)F_{05}$ &  0 & 10 &   4 & 2 & 120 & 1 \\
& $30 \pm 21$ & $50 \pm 33$ & & & seen & \\
$\Lambda^*(1405)S_{01}$ & -- & 0 & -- & -- & & \\
& & $50 \pm 2$ & & & & \\
$\Lambda^*(1520)D_{03}$ & 10 &  28 & -- & -- & & \\
& $7 \pm 1$ & $7 \pm 1$ & & & & \\
$\Lambda^*(2100)G_{07}$ & 18 & 22 &   4 & 2 & & \\
& $53 \pm 24$ & $\sim 9 \pm 4$ & $< 3 \pm 3$ & $< 3 \pm 3$ & & \\
& & & & & & \\
\hline
\end{tabular}
\end{table}

\clearpage

\begin{table}
\centering
\caption[]{As Table~\protect\ref{nuc}, but for $\Xi$ resonances.}
\label{xi} 
\vspace{15pt}
\begin{tabular}{l|cccc|cccc}
\hline
& & & & & & & & \\
Baryon & $\Sigma \overline{K}$ & $\Lambda \overline{K}$ & $\Xi \pi$ 
& $\Xi \eta$ & $\Sigma^* \overline{K}$ & $\Xi^* \pi$ & $\Xi^* \eta$ 
& $\Omega K$ \\
& & & & & & & & \\
\hline
& & & & & & & & \\
$\Xi(1820)D_{13}$ & 30 & 18 &  6 & -- & -- &  3 & -- & -- \\
& $7 \pm 4$ & $7 \pm 4$ & $2 \pm 2$  & & & $7 \pm 4$ & & \\
$\Xi^*(1530)P_{13}$ & -- & -- & 22 & -- & -- & -- & -- & -- \\
& & & $10 \pm 1$ & & & & & \\
& & & & & & & & \\
\hline
\end{tabular}
\end{table}

\clearpage

\begin{table}
\centering
\caption[]{Strong decay widths of missing nucleon resonances 
in MeV. Tentative assignments of one and two star 
resonances are labeled by $^{\ddagger}$.} 
\label{strong1} 
\vspace{15pt} 
\small 
\begin{tabular}{ccl|rrrr|rr}
\hline
& & & & & & & & \\
$N$ & ($v_1,v_2$) & Mass & $N \pi$ & $N \eta$ & $\Sigma K$ 
& $\Lambda K$ & $\Delta \pi$ & $\Sigma^* K$ \\
& & & & & & & & \\
\hline
& & & & & & & & \\
$^{2}8_{J}  [20,1^+]$ & (0,0) & 1713 
&   0 &  0 &  0 &  0 &   0 & -- \\
$^{2}8_{3/2}[70,2^+]$ & (0,0) & 1874 $^{\ddagger}$ 
&  56 &  9 &  0 &  3 &  56 & -- \\
$^{2}8_{5/2}[70,2^+]$ & (0,0) & 1874 
&  84 & 19 &  0 &  8 &  43 & -- \\
$^{2}8_{J}  [70,2^-]$ & (0,0) & 1874 
&   0 &  0 &  0 &  0 &   0 & -- \\
$^{4}8_{1/2}[70,2^+]$ & (0,0) & 1975 
&  19 & 18 &  7 &  0 &  16 &  0 \\
$^{4}8_{3/2}[70,2^+]$ & (0,0) 
& 1975 &  10 &  9 &  4 &  0 &  96 &  0 \\
$^{4}8_{5/2}[70,2^+]$ & (0,0) & 1975 $^{\ddagger}$ 
&   4 &  5 &  3 &  0 & 159 &  0 \\
$^{4}8_{7/2}[70,2^+]$ & (0,0) & 1975 $^{\ddagger}$ 
&  18 & 20 & 12 &  0 &  98 &  0 \\
$^{4}8_{J}  [70,2^-]$ & (0,0) & 1975 
&   0 &  0 &  0 &  0 &   0 &  0 \\
& & & & & & & & \\
$^{2}8_{1/2}[70,1^-]$ & (1,0) & 1909 
& 22 & 1 & 0 & 0 & 1 & 0 \\
$^{2}8_{3/2}[70,1^-]$ & (1,0) & 1909 
& 28 & 1 & 0 & 0 & 1 & 0 \\
$^{4}8_{1/2}[70,1^-]$ & (1,0) & 2009 
& 10 & 4 & 0 & 0 & 3 & 0 \\
$^{4}8_{3/2}[70,1^-]$ & (1,0) & 2009 
& 1 & 1 & 0 & 0 & 21 & 2 \\
$^{4}8_{5/2}[70,1^-]$ & (1,0) & 2009 
& 8 & 4 & 0 & 0 & 13 & 2 \\
$^{2}8_{ J }[20,1^+]$ & (1,0) & 2034 
& 0 & 0 & 0 & 0 & 0 & 0 \\
& & & & & & & & \\
$^{4}8_{3/2}[70,0^+]$ & (0,1) & 1796 
& 14 & 10 & 1 & 0 & 91 & -- \\
$^{2}8_{1/2}[56,1^-]$ & (0,1) & 1847 
& 96 & 2 & 0 & 1 & 35 & -- \\
$^{2}8_{3/2}[56,1^-]$ & (0,1) & 1847 
& 123 & 3 & 0 & 3 & 30 & -- \\
$^{2}8_{1/2}[70,1^-]$ & (0,1) & 1975 
& 2 & 3 & 0 & 4 & 18 & 2 \\
$^{2}8_{3/2}[70,1^-]$ & (0,1) & 1975 
& 2 & 4 & 1 & 6 & 16 & 1 \\
$^{2}8_{ J }[70,1^+]$ & (0,1) & 1975 
& 0 & 0 & 0 & 0 & 0 & 0 \\
$^{4}8_{1/2}[70,1^-]$ & (0,1) & 2072 
& 0 & 2 & 7 & 0 & 3 & 1 \\ 
$^{4}8_{3/2}[70,1^-]$ & (0,1) & 2072 
& 0 & 0 & 1 & 0 & 22 & 6 \\ 
$^{4}8_{5/2}[70,1^-]$ & (0,1) & 2072 
& 0 & 1 & 6 & 0 & 13 & 5 \\ 
$^{4}8_{ J }[70,1^+]$ & (0,1) & 2072 
& 0 & 0 & 0 & 0 & 0 & 0 \\
$^{2}8_{ J }[20,1^-]$ & (0,1) & 2096 
& 0 & 0 & 0 & 0 & 0 & 0 \\
& & & & & & & & \\
\hline
\end{tabular}
\normalsize
\end{table}

\clearpage

\begin{table}
\centering
\caption[]{As Table~\protect\ref{strong1}, but for missing  
$\Sigma$ resonances.} 
\label{strong2} 
\vspace{15pt} 
\small
\begin{tabular}{ccl|rrrrr|rrrr}
\hline
& & & & & & & & & & & \\
$\Sigma$ & ($v_1,v_2$) & Mass & $N \overline{K}$ & $\Sigma \pi$ 
& $\Lambda \pi$ & $\Sigma \eta$ & $\Xi K$ & $\Delta \overline{K}$ 
& $\Sigma^* \pi$ & $\Sigma^* \eta$ & $\Xi^* K$ \\
& & & & & & & & & & & \\
\hline
& & & & & & & & & & & \\
$^{4}8_{1/2}[70,1^-]$ & (0,0) & 1822 
& 78 &  20 & 39 &  0 &  0 &  5 &  5 & --  & -- \\
$^{4}8_{3/2}[70,1^-]$ & (0,0) & 1822 
& 12 &   3 &  5 &  0 &  0 & 20 & 31 & -- & -- \\
$^{2}8_{J}  [20,1^+]$ & (0,0) & 1849 
&  0 &   0 &  0 &  0 &  0 &  0 &  0 & -- & -- \\
$^{2}8_{3/2}[56,2^+]$ & (0,0) & 1872 
&  1 &   9 &  5 &  0 &  0 &  2 &  2 & -- & -- \\
$^{2}8_{3/2}[70,2^+]$ & (0,0) & 1999 
&  2 &  43 &  3 &  1 &  1 & 21 &  9 &  0 & -- \\
$^{2}8_{5/2}[70,2^+]$ & (0,0) & 1999 
&  3 &  68 &  5 &  4 &  4 & 14 &  7 &  0 & -- \\
$^{2}8_{J}  [70,2^-]$ & (0,0) & 1999 
&  0 &   0 &  0 &  0 &  0 &  0 &  0 &  0 & -- \\
$^{4}8_{1/2}[70,2^+]$ & (0,0) & 2095 
& 43 &  10 & 19 &  1 &  1 &  6 &  2 &  0 &  0 \\
$^{4}8_{3/2}[70,2^+]$ & (0,0) & 2095 
& 22 &   5 &  9 &  0 &  1 & 41 & 15 &  0 &  0 \\
$^{4}8_{5/2}[70,2^+]$ & (0,0) & 2095 
& 10 &   2 &  4 &  0 &  0 & 70 & 25 &  1 &  0 \\
$^{4}8_{7/2}[70,2^+]$ & (0,0) & 2095 
& 43 &  10 & 18 &  1 &  2 & 47 & 15 &  1 &  0 \\
$^{4}8_{J}  [70,2^-]$ & (0,0) & 2095 
&  0 &   0 &  0 &  0 &  0  &  0 &  0 &  0 &  0 \\
& & & & & & & & & & & \\
$^{2}8_{1/2}[70,1^-]$ & (1,0) & 2033 
& 1 & 10 & 1 & 0 & 1 & 0 & 0 & 0 & 0 \\
$^{2}8_{3/2}[70,1^-]$ & (1,0) & 2033 
& 1 & 14 & 2 & 0 & 2 & 0 & 0 & 0 & 0 \\
$^{4}8_{1/2}[70,1^-]$ & (1,0) & 2127 
& 22 & 4 & 9 & 0 & 0 & 1 & 0 & 0 & 0 \\
$^{4}8_{3/2}[70,1^-]$ & (1,0) & 2127 
& 3 & 1 & 1 & 0 & 0 & 3 & 2 & 1 & 1 \\
$^{4}8_{5/2}[70,1^-]$ & (1,0) & 2127 
& 18 & 3 & 7 & 0 & 0 & 2 & 2 & 0 & 1 \\
$^{2}8_{ J }[20,1^+]$ & (1,0) & 2150 
& 0 & 0 & 0 & 0 & 0 & 0 & 0 & 0 & 0 \\
& & & & & & & & & & & \\
$^{2}8_{1/2}[70,0^+]$ & (0,1) & 1822 $^{\ddagger}$ 
& 2 & 56 & 5 & 0 & 0 & 1 & 4 & -- & -- \\
$^{4}8_{3/2}[70,0^+]$ & (0,1) & 1926 
& 31 & 7 & 14 & 0 & 0 & 23 & 14 & -- & -- \\
$^{2}8_{1/2}[56,1^-]$ & (0,1) & 1974 
& 2 & 27 & 15 & 1 & 0 & 9 & 5 & 0 & -- \\
$^{2}8_{1/2}[70,1^-]$ & (0,1) & 2095 
& 0 & 4 & 0 & 2 & 5 & 13 & 3 & 1 & 1 \\
$^{2}8_{3/2}[70,1^-]$ & (0,1) & 2095 
& 0 & 4 & 0 & 3 & 7 & 11 & 3 & 1 & 0 \\
$^{2}8_{ J }[70,1^+]$ & (0,1) & 2095 
& 0 & 0 & 0 & 0 & 0 & 0 & 0 & 0 & 0 \\
$^{4}8_{1/2}[70,1^-]$ & (0,1) & 2186 
& 0 & 0 & 0 & 0 & 1 & 2 & 1 & 0 & 1 \\
$^{4}8_{3/2}[70,1^-]$ & (0,1) & 2186 
& 0 & 0 & 0 & 0 & 0 & 17 & 4 & 2 & 3 \\
$^{4}8_{5/2}[70,1^-]$ & (0,1) & 2186 
& 0 & 0 & 0 & 0 & 1 & 10 & 2 & 1 & 3 \\
$^{4}8_{ J }[70,1^+]$ & (0,1) & 2186 
& 0 & 0 & 0 & 0 & 0 & 0 & 0 & 0 & 0 \\
$^{2}8_{ J }[20,1^-]$ & (0,1) & 2209 
& 0 & 0 & 0 & 0 & 0 & 0 & 0 & 0 & 0 \\
& & & & & & & & & & & \\
\hline
\end{tabular}
\normalsize
\end{table}

\clearpage

\begin{table}
\centering
\caption[]{As Table~\protect\ref{strong1}, but for missing 
$\Lambda$ resonances.} 
\label{strong3} 
\vspace{15pt} 
\small
\begin{tabular}{ccl|rrrr|rr}
\hline
& & & & & & & & \\
$\Lambda$ & ($v_1,v_2$) & Mass & $N \overline{K}$ & $\Sigma \pi$ 
& $\Lambda \eta$ & $\Xi K$ & $\Sigma^* \pi$ & $\Xi^* K$ \\
& & & & & & & & \\
\hline
& & & & & & & & \\
$^{4}8_{3/2}[70,1^-]$ & (0,0) & 1799 
&   0 & 11 &  2 & -- & 112 & -- \\
$^{2}8_{J}  [20,1^+]$ & (0,0) & 1826 
&   0 &  0 &  0 &  0 &   0 & -- \\
$^{2}8_{3/2}[70,2^+]$ & (0,0) & 1978 
&  46 &  7 &  0 &  0 &  33 & -- \\
$^{2}8_{5/2}[70,2^+]$ & (0,0) & 1978 
&  77 & 11 &  0 &  2 &  25 & -- \\
$^{2}8_{J}  [70,2^-]$ & (0,0) & 1978 
&   0 &  0 &  0 &  0 &   0 & -- \\
$^{4}8_{1/2}[70,2^+]$ & (0,0) & 2074 
&   0 & 43 & 13 &  3 &  9 &  0 \\
$^{4}8_{3/2}[70,2^+]$ & (0,0) & 2074 
&   0 & 21 &  7 &  2 &  59 &  0 \\
$^{4}8_{7/2}[70,2^+]$ & (0,0) & 2074 
&   0 & 42 & 17 &  5 &  62 &  0 \\
$^{4}8_{J}  [70,2^-]$ & (0,0) & 2074 
&   0 &  0 &  0 &  0 &   0 &  0 \\
& & & & & & & & \\
$^{2}8_{1/2}[70,1^-]$ & (1,0) & 2012 
& 14 & 1 & 0 & 1 & 0 & -- \\
$^{2}8_{3/2}[70,1^-]$ & (1,0) & 2012 
& 20 & 2 & 0 & 2 & 0 & -- \\
$^{4}8_{1/2}[70,1^-]$ & (1,0) & 2107 
& 0 & 16 & 2 & 0 & 1 & 1 \\
$^{4}8_{3/2}[70,1^-]$ & (1,0) & 2107 
& 0 & 2 & 0 & 0 & 7 & 2 \\
$^{4}8_{5/2}[70,1^-]$ & (1,0) & 2107 
& 0 & 12 & 2 & 0 & 4 & 2 \\
$^{2}8_{ J }[20,1^+]$ & (1,0) & 2130 
& 0 & 0 & 0 & 0 & 0 & 0 \\
& & & & & & & & \\
$^{4}8_{3/2}[70,0^+]$ & (0,1) & 1904 
& 0 & 30 & 7 & 0 & 52 & -- \\
$^{2}8_{1/2}[56,1^-]$ & (0,1) & 1952 
& 47 & 27 & 1 & 0 & 19 & -- \\
$^{2}8_{3/2}[56,1^-]$ & (0,1) & 1952 
& 66 & 36 & 1 & 0 & 16 & -- \\
$^{2}8_{1/2}[70,1^-]$ & (0,1) & 2074 
& 3 & 1 & 0 & 3 & 15 & 1 \\
$^{2}8_{3/2}[70,1^-]$ & (0,1) & 2074 
& 3 & 1 & 0 & 5 & 13 & 0 \\
$^{2}8_{ J }[70,1^+]$ & (0,1) & 2074 
& 0 & 0 & 0 & 0 & 0 & 0 \\
$^{4}8_{1/2}[70,1^-]$ & (0,1) & 2167 
& 0 & 1 & 2 & 4 & 3 & 2 \\
$^{4}8_{3/2}[70,1^-]$ & (0,1) & 2167 
& 0 & 0 & 0 & 1 & 20 & 8 \\
$^{4}8_{5/2}[70,1^-]$ & (0,1) & 2167 
& 0 & 1 & 2 & 3 & 12 & 7 \\
$^{4}8_{ J }[70,1^+]$ & (0,1) & 2167 
& 0 & 0 & 0 & 0 & 0 & 0 \\
$^{2}8_{ J }[20,1^-]$ & (0,1) & 2190 
& 0 & 0 & 0 & 0 & 0 & 0 \\
& & & & & & & & \\
\hline
\end{tabular}
\normalsize
\end{table}

\clearpage

\begin{table}
\centering
\caption[]{As Table~\protect\ref{strong1}, but for missing 
$\Xi$ resonances.} 
\label{strong4} 
\vspace{15pt} 
\small
\begin{tabular}{ccl|rrrr|rrrr}
\hline
& & & & & & & & & & \\
$\Xi$ & ($v_1,v_2$) & Mass & $\Sigma \overline{K}$ 
& $\Lambda \overline{K}$ & $\Xi \pi$ & $\Xi \eta$ 
& $\Sigma^* \overline{K}$ & $\Xi^* \pi$ & $\Xi^* \eta$ 
& $\Omega K$ \\
& & & & & & & & & & \\
\hline
& & & & & & & & & & \\
$^{2}8_{1/2}[70,1^-]$ & (0,0) & 1828 
&  11 & 10 &  4 & -- & -- &  6 & -- & -- \\
$^{4}8_{1/2}[70,1^-]$ & (0,0) & 1932 
&  14 & 24 & 119 &  0 &  1 &  6 & -- & -- \\
$^{4}8_{3/2}[70,1^-]$ & (0,0) & 1932 
&   3 &  4 & 17 &  0 &  2 & 34 & -- & -- \\
$^{4}8_{5/2}[70,1^-]$ & (0,0) & 1932 
&  15 & 23 & 100 &  0 &  2 & 24 & -- & -- \\
$^{2}8_{J}  [20,1^+]$ & (0,0) & 1957 
&   0 &  0 &   0 &  0 &  0 &  0 & -- & -- \\
$^{2}8_{3/2}[56,2^+]$ & (0,0) & 1979 
&   8 &  1 &  1 &  0 &  0 &  2 & -- & -- \\
$^{2}8_{5/2}[56,2^+]$ & (0,0) & 1979 
&  20 &  1 &  1 &  0 &  0 &  1 & -- & -- \\
$^{2}8_{3/2}[70,2^+]$ & (0,0) & 2100 
&  25 &  9 &  3 &  1 &  5 & 10 &  0 & -- \\
$^{2}8_{5/2}[70,2^+]$ & (0,0) & 2100 
&  47 & 16 &  4 &  2 &  3 &  7 &  0 & -- \\
$^{2}8_{J}  [70,2^-]$ & (0,0) & 2100 
&   0 &  0 &  0 &  0 &  0 &  0 &  0 & -- \\
$^{4}8_{1/2}[70,2^+]$ & (0,0) & 2191 
&  11 & 14 & 60 &  1 &  2 &  3 &  0 &  0 \\
$^{4}8_{3/2}[70,2^+]$ & (0,0) & 2191 
&   5 &  7 & 30 &  1 & 11 & 18 &  0 &  0 \\
$^{4}8_{5/2}[70,2^+]$ & (0,0) & 2191 
&   3 &  3 & 13 &  0 & 19 & 30 &  0 &  0 \\
$^{4}8_{7/2}[70,2^+]$ & (0,0) & 2191 
&  12 & 15 & 59 &  2 & 13 & 19 &  0 &  0 \\
$^{4}8_{J}  [70,2^-]$ & (0,0) & 2191 
&   0 &  0 &  0 &  0 &  0 &  0 &  0 &  0 \\
& & & & & & & & & & \\
$^{2}8_{1/2}[56,0^+]$ & (1,0) & 1727 
& 0 & 0 & 2 & -- & -- & 0 & -- & -- \\
$^{2}8_{1/2}[70,1^-]$ & (1,0) & 2132 
& 3 & 2 & 1 & 0 & 1 & 0 & 0 & -- \\
$^{2}8_{3/2}[70,1^-]$ & (1,0) & 2132 
& 5 & 3 & 1 & 0 & 1 & 0 & 0 & -- \\
$^{4}8_{1/2}[70,1^-]$ & (1,0) & 2222 
& 3 & 6 & 22 & 0 & 0 & 0 & 0 & 0 \\
$^{4}8_{3/2}[70,1^-]$ & (1,0) & 2222 
& 0 & 1 & 3 & 0 & 0 & 1 & 1 & 1 \\
$^{4}8_{5/2}[70,1^-]$ & (1,0) & 2222 
& 3 & 5 & 17 & 0 & 0 & 1 & 0 & 1 \\
$^{2}8_{ J }[20,1^+]$ & (1,0) & 2244 
& 0 & 0 & 0 & 0 & 0 & 0 & 0 & 0 \\
& & & & & & & & & & \\
$^{2}8_{1/2}[70,0^+]$ & (0,1) & 1932 
& 24 & 11 & 3 & 0 & 0 & 4 & -- & -- \\
$^{4}8_{3/2}[70,0^+]$ & (0,1) & 2030 
& 7 & 10 & 44 & 0 & 4 & 15 & -- & -- \\
$^{2}8_{1/2}[56,1^-]$ & (0,1) & 2076 
& 34 & 2 & 2 & 1 & 2 & 5 & -- & -- \\
$^{2}8_{3/2}[56,1^-]$ & (0,1) & 2076 
& 52 & 3 & 3 & 1 & 1 & 4 & -- & -- \\
$^{2}8_{1/2}[70,1^-]$ & (0,1) & 2191 
& 5 & 1 & 0 & 1 & 6 & 5 & 1 & 0 \\
$^{2}8_{3/2}[70,1^-]$ & (0,1) & 2191 
& 6 & 1 & 0 & 2 & 5 & 5 & 0 & 0 \\
$^{2}8_{ J }[70,1^+]$ & (0,1) & 2191 
& 0 & 0 & 0 & 0 & 0 & 0 & 0 & 0 \\
$^{4}8_{1/2}[70,1^-]$ & (0,1) & 2278 
& 1 & 0 & 2 & 0 & 1 & 1 & 0 & 1 \\
$^{4}8_{3/2}[70,1^-]$ & (0,1) & 2278 
& 0 & 0 & 0 & 0 & 8 & 8 & 2 & 6 \\
$^{4}8_{5/2}[70,1^-]$ & (0,1) & 2278 
& 0 & 0 & 1 & 0 & 5 & 4 & 1 & 5 \\
$^{4}8_{ J }[70,1^+]$ & (0,1) & 2278 
& 0 & 0 & 0 & 0 & 0 & 0 & 0 & 0 \\
$^{2}8_{ J }[20,1^-]$ & (0,1) & 2300 
& 0 & 0 & 0 & 0 & 0 & 0 & 0 & 0 \\
& & & & & & & & & & \\
\hline
\end{tabular}
\normalsize
\end{table}

\clearpage

\begin{table}
\centering
\caption[]{As Table~\protect\ref{strong1}, but for missing 
$\Delta$ resonances.} 
\label{strong5} 
\vspace{15pt} 
\begin{tabular}{ccl|rr|rrr}
\hline
& & & & & & & \\
$\Delta$ & ($v_1,v_2$) & Mass & $N \pi$ & $\Sigma K$ 
& $\Delta \pi$ & $\Delta \eta$ & $\Sigma^* K$ \\
& & & & & & & \\
\hline
& & & & & & & \\
$^{2}10_{3/2}[70,2^+]$ & (0,0) & 1946 
&   9 &  1 & 106 &  5 &  0 \\
$^{2}10_{5/2}[70,2^+]$ & (0,0) & 1946 $^{\ddagger}$ 
&  13 &  2 &  84 &  3 &  0 \\
$^{2}10_{3/2}[70,2^-]$ & (0,0) & 1946 
&   0 &  0 &   0 &  0 &  0 \\
& & & & & & & \\
$^{2}10_{1/2}[70,1^-]$ & (1,0) & 1981 $^{\ddagger}$ 
& 4 & 0 & 9 & 3 & 3 \\
$^{2}10_{3/2}[70,1^-]$ & (1,0) & 1981 
& 6 & 0 & 8 & 3 & 2 \\
& & & & & & & \\
$^{2}10_{1/2}[70,0^+]$ & (0,1) & 1764 $^{\ddagger}$ 
& 13 & 0 & 71 & -- & -- \\
$^{4}10_{1/2}[56,1^-]$ & (0,1) & 2020 
& 113 & 12 & 18 & 1 & 0 \\
$^{4}10_{3/2}[56,1^-]$ & (0,1) & 2020 
& 14 & 2 & 128 & 6 & 0 \\
$^{4}10_{5/2}[56,1^-]$ & (0,1) & 2020 
& 84 & 12 & 77 & 4 & 0 \\
$^{2}10_{1/2}[70,1^-]$ & (0,1) & 2044 
& 0 & 1 & 17 & 12 & 8 \\
$^{2}10_{3/2}[70,1^-]$ & (0,1) & 2044 
& 0 & 1 & 15 & 9 & 5 \\
$^{2}10_{ J }[70,1^+]$ & (0,1) & 2044 
& 0 & 0 & 0 & 0 & 0 \\
& & & & & & & \\
\hline
\end{tabular}
\end{table}

\clearpage

\begin{table}
\centering
\caption[]{As Table~\protect\ref{strong1}, but for missing 
$\Sigma^*$ resonances.} 
\label{strong6} 
\vspace{15pt} 
\begin{tabular}{ccl|rrrrr|rrrr}
\hline
& & & & & & & & & & & \\
$\Sigma^*$ & ($v_1,v_2$) & Mass & $N \overline{K}$ & $\Sigma \pi$ 
& $\Lambda \pi$ & $\Sigma \eta$ & $\Xi K$ & $\Delta \overline{K}$ 
& $\Sigma^* \pi$ & $\Sigma^* \eta$ & $\Xi^* K$ \\
& & & & & & & & & & & \\
\hline
& & & & & & & & & & & \\
$^{2}10_{1/2}[70,1^-]$ & (0,0) & 1755 
&  3 &  4 &  7 &  0 & -- &  1 & 42 & -- & -- \\
$^{2}10_{3/2}[70,1^-]$ & (0,0) & 1755 
&  5 &  5 & 10 &  0 & -- &  1 & 32 & -- & -- \\
$^{4}10_{1/2}[56,2^+]$ & (0,0) & 2012 
& 11 & 10 & 19 &  2 &  0 &  1 &  2 &  0 & -- \\
$^{4}10_{3/2}[56,2^+]$ & (0,0) & 2012 
&  5 &  5 &  9 &  1 &  0 &  7 & 12 &  0 & -- \\
$^{4}10_{5/2}[56,2^+]$ & (0,0) & 2012 
&  2 &  2 &  4 &  1 &  0 & 11 & 20 &  0 & -- \\
$^{2}10_{3/2}[70,2^+]$ & (0,0) & 2037 
&  2 &  2 &  4 &  1 &  0 & 30 & 44 &  0 &  0 \\
$^{2}10_{5/2}[70,2^+]$ & (0,0) & 2037 
&  4 &  3 &  6 &  1 &  0 & 21 & 34 &  0 &  0 \\
$^{2}10_{J}  [70,2^-]$ & (0,0) & 2037 
&  0 &  0 &  0 &  0 &  0 &  0 &  0 &  0 &  0 \\
& & & & & & & & & & & \\
$^{4}10_{3/2}[56,0^+]$ & (1,0) & 1765 
& 28 & 26 & 58 & 0 & -- & 0 & 18 & -- & -- \\
$^{2}10_{1/2}[70,1^-]$ & (1,0) & 2070 
& 1 & 1 & 2 & 0 & 0 & 0 & 1 & 1 & 1 \\
$^{2}10_{3/2}[70,1^-]$ & (1,0) & 2070 
& 1 & 1 & 2 & 0 & 0 & 0 & 1 & 0 & 0 \\
& & & & & & & & & & & \\
$^{2}10_{1/2}[70,0^+]$ & (0,1) & 1863 
& 3 & 3 & 6 & 0 & 0 & 5 & 25 & -- & -- \\
$^{4}10_{1/2}[56,1^-]$ & (0,1) & 2107 
& 29 & 26 & 50 & 10 & 3 & 5 & 7 & 0 & 0 \\
$^{4}10_{3/2}[56,1^-]$ & (0,1) & 2107 
& 4 & 3 & 6 & 2 & 0 & 35 & 51 & 0 & 0 \\
$^{4}10_{5/2}[56,1^-]$ & (0,1) & 2107 
& 23 & 20 & 37 & 9 & 3 & 23 & 31 & 0 & 0 \\
$^{2}10_{1/2}[70,1^-]$ & (0,1) & 2131 
& 0 & 0 & 0 & 0 & 0 & 12 & 12 & 1 & 5 \\
$^{2}10_{3/2}[70,1^-]$ & (0,1) & 2131 
& 0 & 0 & 0 & 1 & 0 & 11 & 10 & 1 & 3 \\
$^{2}10_{ J }[70,1^+]$ & (0,1) & 2131 
& 0 & 0 & 0 & 0 & 0 & 0 & 0 & 0 & 0 \\
& & & & & & & & & & & \\
\hline
\end{tabular}
\end{table}

\clearpage

\begin{table}
\centering
\caption[]{As Table~\protect\ref{strong1}, but for missing 
$\Xi^*$ resonances.} 
\label{strong7}
\vspace{15pt} 
\begin{tabular}{ccl|rrrr|rrrr}
\hline
& & & & & & & & & & \\
$\Xi^*$ & ($v_1,v_2$) & Mass & $\Sigma \overline{K}$ 
& $\Lambda \overline{K}$ & $\Xi \pi$ & $\Xi \eta$ 
& $\Sigma^* \overline{K}$ & $\Xi^* \pi$ & $\Xi^* \eta$ 
& $\Omega K$ \\
& & & & & & & & & & \\
\hline
& & & & & & & & & & \\
$^{2}10_{1/2}[70,1^-]$ & (0,0) & 1869 
&  2 &  4 &  5 &  0  & -- & 11 & -- & -- \\
$^{2}10_{3/2}[70,1^-]$ & (0,0) & 1869 
&  4 &  7 &  8 &  0  & -- &  8 & -- & -- \\
$^{4}10_{1/2}[56,2^+]$ & (0,0) & 2112 
&  9 & 13 & 14 &  2 &  1 &  1 &  0 & -- \\
$^{4}10_{3/2}[56,2^+]$ & (0,0) & 2112 
&  4 &  7 &  7 &  1 &  6 &  3 &  0 & -- \\
$^{4}10_{5/2}[56,2^+]$ & (0,0) & 2112 
&  2 &  3 &  3 &  1 & 10 &  6 &  0 & -- \\
$^{4}10_{7/2}[56,2^+]$ & (0,0) & 2112 
& 11 & 15 & 14 &  3 &  8 &  4 &  0 & -- \\
$^{2}10_{3/2}[70,2^+]$ & (0,0) & 2135 
&  2 &  3 &  3 &  0 & 30 & 13 &  0 & -- \\
$^{2}10_{5/2}[70,2^+]$ & (0,0) & 2135 
&  4 &  5 &  5 &  1 & 20 & 10 &  0 & -- \\
$^{2}10_{J}  [70,2^-]$ & (0,0) & 2135 
&  0 &  0 &  0 &  0  &  0 &  0 &  0 &  0 \\
& & & & & & & & & & \\
$^{4}10_{3/2}[56,0^+]$ & (1,0) & 1878 
& 12 & 29 & 38 & 0 & -- & 4 & -- & -- \\ 
$^{2}10_{1/2}[70,1^-]$ & (1,0) & 2167 
& 0 & 1 & 1 & 0 & 1 & 0 & 0 & -- \\
$^{2}10_{3/2}[70,1^-]$ & (1,0) & 2167 
& 1 & 1 & 1 & 0 & 1 & 0 & 0 & -- \\
& & & & & & & & & & \\
$^{2}10_{1/2}[70,0^+]$ & (0,1) & 1970 
& 2 & 4 & 4 & 0 & 2 & 6 & -- & -- \\
$^{4}10_{1/2}[56,1^-]$ & (0,1) & 2203 
& 28 & 38 & 39 & 8 & 5 & 2 & 0 & 0 \\
$^{4}10_{3/2}[56,1^-]$ & (0,1) & 2203 
& 4 & 5 & 5 & 1 & 34 & 15 & 0 & 0 \\
$^{4}10_{5/2}[56,1^-]$ & (0,1) & 2203 
& 23 & 30 & 30 & 8 & 23 & 9 & 0 & 0 \\
$^{2}10_{1/2}[70,1^-]$ & (0,1) & 2225 
& 0 & 0 & 0 & 0 & 21 & 5 & 0 & 1 \\
$^{2}10_{3/2}[70,1^-]$ & (0,1) & 2225 
& 0 & 0 & 0 & 1 & 18 & 4 & 0 & 1 \\
$^{2}10_{ J }[70,1^+]$ & (0,1) & 2225 
& 0 & 0 & 0 & 0 & 0 & 0 & 0 & 0 \\
& & & & & & & & & & \\
\hline
\end{tabular}
\end{table}

\clearpage

\begin{table}
\centering
\caption[]{As Table~\protect\ref{strong1}, but for missing 
$\Omega$ resonances.} 
\label{strong8} 
\vspace{15pt} 
\begin{tabular}{ccl|r|rr}
\hline
& & & & & \\
$\Omega$ & ($v_1,v_2$) & Mass & $\Xi \overline{K}$ 
& $\Xi^* \overline{K}$ & $\Omega \eta$ \\
& & & & & \\
\hline
& & & & & \\
$^{2}10_{1/2}[70,1^-]$ & (0,0) & 1989 &  7 & -- & -- \\
$^{2}10_{3/2}[70,1^-]$ & (0,0) & 1989 & 15 & -- & -- \\
$^{4}10_{1/2}[56,2^+]$ & (0,0) & 2219 & 35 &  0 & -- \\
$^{4}10_{3/2}[56,2^+]$ & (0,0) & 2219 & 18 &  3 & -- \\
$^{4}10_{5/2}[56,2^+]$ & (0,0) & 2219 & 10 &  6 & -- \\
$^{4}10_{7/2}[56,2^+]$ & (0,0) & 2219 & 43 &  5 & -- \\
$^{2}10_{3/2}[70,2^+]$ & (0,0) & 2242 &  8 & 20 &  0 \\
$^{2}10_{5/2}[70,2^+]$ & (0,0) & 2242 & 14 & 12 &  0 \\
$^{2}10_{J}  [70,2^-]$ & (0,0) & 2242 &  0 &  0 &  0 \\
& & & & & \\
$^{4}10_{3/2}[56,0^+]$ & (1,0) & 1998 & 49 & -- & -- \\
$^{2}10_{1/2}[70,1^-]$ & (1,0) & 2272 & 1 & 3 & 1 \\
$^{2}10_{3/2}[70,1^-]$ & (1,0) & 2272 & 2 & 3 & 0 \\
& & & & & \\
$^{2}10_{1/2}[70,0^+]$ & (0,1) & 2085 & 9 & 0 & -- \\
$^{4}10_{1/2}[56,1^-]$ & (0,1) & 2306 & 109 & 4 & 0 \\
$^{4}10_{3/2}[56,1^-]$ & (0,1) & 2306 & 15 & 22 & 0 \\
$^{4}10_{5/2}[56,1^-]$ & (0,1) & 2306 & 90 & 15 & 0 \\
$^{2}10_{1/2}[70,1^-]$ & (0,1) & 2327 & 1 & 23 & 4 \\
$^{2}10_{3/2}[70,1^-]$ & (0,1) & 2327 & 1 & 19 & 3 \\
$^{2}10_{ J }[70,1^+]$ & (0,1) & 2327 & 0 & 0 & 0 \\
& & & & & \\
\hline
\end{tabular}
\end{table}

\clearpage

\begin{table}
\centering
\caption[]{As Table~\protect\ref{strong1}, but for missing 
$\Lambda^*$ resonances.} 
\label{strong9} 
\vspace{15pt} 
\begin{tabular}{ccl|rrrr}
\hline
& & & & & & \\
$\Lambda^*$ & ($v_1,v_2$) & Mass & $N \overline{K}$ & $\Sigma \pi$ 
& $\Lambda \eta$ & $\Xi K$ \\ 
& & & & & & \\
\hline
& & & & & & \\
$^{4}1_{J}  [20,1^+]$ & (0,0) & 1891 &  0 &   0 &  0 &  0 \\
$^{2}1_{3/2}[70,2^+]$ & (0,0) & 1939 & 38 &  52 &  4 &  0 \\
$^{2}1_{5/2}[70,2^+]$ & (0,0) & 1939 & 66 &  85 & 10 &  2 \\
$^{2}1_{J}  [70,2^-]$ & (0,0) & 1939 &  0 &   0 &  0 &  0 \\
& & & & & & \\
$^{2}1_{1/2}[70,1^-]$ & (1,0) & 1974 & 9 & 7 & 0 & 2 \\ 
$^{2}1_{3/2}[70,1^-]$ & (1,0) & 1974 & 13 & 10 & 0 & 4 \\ 
$^{4}1_{J}  [20,1^+]$ & (1,0) & 2186 & 0 & 0 & 0 & 0 \\
& & & & & & \\
$^{2}1_{1/2}[70,0^+]$ & (0,1) & 1756 & 43 & 59 & 0 & -- \\
$^{2}1_{1/2}[70,1^-]$ & (0,1) & 2038 & 3 & 8 & 4 & 6 \\
$^{2}1_{3/2}[70,1^-]$ & (0,1) & 2038 & 4 & 10 & 5 & 11 \\
$^{2}1_{J}  [70,1^+]$ & (0,1) & 2038 & 0 & 0 & 0 & 0 \\
$^{4}1_{J}  [20,1^-]$ & (0,1) & 2244 & 0 & 0 & 0 & 0 \\
& & & & & & \\
\hline
\end{tabular}
\end{table}

\clearpage

\begin{table}
\centering
\caption[]{The spin-flip amplitudes of Eq.~(\ref{ab}), associated
with transverse helicity $\nu=1/2$ amplitudes for 
$^{2}8_{1/2}[56,0^+] \, \rightarrow \,^{2}8_{1/2}[56,0^+] + \gamma$ 
and $^{4}10_{3/2}[56,0^+] \, \rightarrow \,^{2}8_{1/2}[56,0^+] 
+ \gamma$ couplings with ${\cal F}(k)=1/(1+k^2a^2)^2$ 
from Table~\ref{collff}. 
The orbit-flip amplitudes are ${\cal A}_{3/2}={\cal A}_{1/2}=0$.}
\label{gamma1}
\vspace{15pt}
\begin{tabular}{lcc}
\hline
& & \\
\mbox{Coupling} & ${\cal B}_{1/2}$ & ${\cal B}_{3/2}$ \\
& & \\
\hline
& & \\
$\Sigma^0 \rightarrow \Lambda + \gamma$ 
& $\frac{1}{3\sqrt{3}} \, \mu_p {\cal F}(k)$ & 0 \\ 
& & \\
$\Delta^+ \rightarrow p + \gamma$ 
& $-\frac{\sqrt{2}}{9} \, \mu_p {\cal F}(k)$ 
& $-\frac{\sqrt{2}}{3\sqrt{3}} \, \mu_p {\cal F}(k)$ \\
$\Delta^0 \rightarrow n + \gamma$ 
& $-\frac{\sqrt{2}}{9} \, \mu_p {\cal F}(k)$ 
& $-\frac{\sqrt{2}}{3\sqrt{3}} \, \mu_p {\cal F}(k)$ \\
$\Sigma^{*,+} \rightarrow \Sigma^+ + \gamma$ 
& $ \frac{\sqrt{2}}{9} \, \mu_p {\cal F}(k)$ 
& $ \frac{\sqrt{2}}{3\sqrt{3}} \, \mu_p {\cal F}(k)$ \\
$\Sigma^{*,0} \rightarrow \Sigma^0 + \gamma$ 
& $\frac{1}{9\sqrt{2}} \, \mu_p {\cal F}(k)$ 
& $\frac{1}{3\sqrt{6}} \, \mu_p {\cal F}(k)$ \\
$\Sigma^{*,0} \rightarrow \Lambda + \gamma$ 
& $-\frac{1}{3\sqrt{6}} \, \mu_p {\cal F}(k)$ 
& $-\frac{1}{3\sqrt{2}} \, \mu_p {\cal F}(k)$ \\
$\Sigma^{*,-} \rightarrow \Sigma^- + \gamma$ & 0 & 0 \\
$\Xi^{*,0} \rightarrow \Xi^0 + \gamma$ 
& $ \frac{\sqrt{2}}{9} \, \mu_p {\cal F}(k)$ 
& $ \frac{\sqrt{2}}{3\sqrt{3}} \, \mu_p {\cal F}(k)$ \\
$\Xi^{*,-} \rightarrow \Xi^- + \gamma$ & 0 & 0 \\
& & \\
\hline
\end{tabular}
\end{table}

\clearpage

\begin{table}
\centering
\caption[]{Orbit- and spin-flip amplitudes of Eq.~(\ref{ab}), 
associated with transverse helicity $\nu=1/2$ amplitudes for 
$^{2}1_J[70,1^-] \, \rightarrow \,^{2}8_{1/2}[56,0^+] + \gamma$ and 
$^{2}1_J[70,1^-] \, \rightarrow \,^{4}10_{3/2}[56,0^+] + \gamma$ 
couplings with ${\cal F}(k)= i \sqrt{3} ka/(1+k^2a^2)^2$ 
from Table~\ref{collff} and ${\cal G}_{-}$ from Eq.~(\ref{gff}). 
The helicity $\nu=3/2$ amplitudes are 
${\cal A}_{3/2}={\cal A}_{1/2}$ and ${\cal B}_{3/2}=0$.}
\label{gamma2}
\vspace{15pt}
\begin{tabular}{lcc}
\hline
& & \\
\mbox{Coupling} & ${\cal A}_{1/2}$ & ${\cal B}_{1/2}$ \\
& & \\
\hline
& & \\
$\Lambda^* \rightarrow \Lambda + \gamma$ 
& $\frac{1}{6\sqrt{2}} \, \mu_p {\cal G}_{-}(k)/g$ 
& $\frac{1}{6\sqrt{2}} \, \mu_p {\cal F}(k)$ \\
$\Lambda^* \rightarrow \Sigma^0 + \gamma$ 
& $\frac{1}{2\sqrt{6}} \, \mu_p {\cal G}_{-}(k)/g$ 
& $\frac{1}{2\sqrt{6}} \, \mu_p {\cal F}(k)$ \\
& & \\
$\Lambda^* \rightarrow \Sigma^{*,0} + \gamma$ & 0 & 0 \\
& & \\
\hline
\end{tabular}
\end{table}

\clearpage

\begin{table}
\centering
\caption[]{Transition magnetic moments associated with the 
$^{2}8[56] \, \rightarrow \,^{2}8[56] + \gamma$ and 
$^{4}10[56] \, \rightarrow \,^{2}8[56] + \gamma$ couplings 
calculated with ${\cal F}(k)=1$ (I) and 
${\cal F}(k)=1/(1+k^2a^2)^2$ (II).}
\label{trmm}
\vspace{15pt}
\begin{tabular}{lcrr}
\hline
& & & \\
$B \rightarrow B' + \gamma$ 
& $\mu_{BB'}(k)$ & (I) & (II) \\
& & & \\
\hline
& & & \\
$\Sigma^0 \rightarrow \Lambda + \gamma$ 
& $\mu_p {\cal F}(k)/\sqrt{3}$ & $ 1.613$ & $ 1.588$ \\
& & & \\
$\Delta^+ \rightarrow p + \gamma$ 
& $ 2\sqrt{2} \, \mu_p {\cal F}(k)/3$ & $ 2.633$ & $ 2.206$ \\
$\Delta^0 \rightarrow n + \gamma$ 
& $ 2\sqrt{2} \, \mu_p {\cal F}(k)/3$ & $ 2.633$ & $ 2.208$ \\
$\Sigma^{*,+} \rightarrow \Sigma^+ + \gamma$ 
& $-2\sqrt{2} \, \mu_p {\cal F}(k)/3$ & $-2.633$ & $-2.411$ \\
$\Sigma^{*,0} \rightarrow \Sigma^0 + \gamma$ 
& $- \sqrt{2} \, \mu_p {\cal F}(k)/3$ & $-1.317$ & $-1.209$ \\
$\Sigma^{*,0} \rightarrow \Lambda + \gamma$ 
& $\sqrt{2} \, \mu_p {\cal F}(k)/\sqrt{3}$ & $ 2.280$ & $ 1.952$ \\
$\Sigma^{*,-} \rightarrow \Sigma^- + \gamma$ 
& 0 & $ 0.000$ & $ 0.000$ \\
$\Xi^{*,0} \rightarrow \Xi^0 + \gamma$ & 
$-2\sqrt{2} \, \mu_p {\cal F}(k)/3$ & $-2.633$ & $-2.361$ \\
$\Xi^{*,-} \rightarrow \Xi^- + \gamma$ 
& 0 & $ 0.000$ & $ 0.000$ \\
& & & \\
\hline
\end{tabular}
\end{table}

\clearpage

\begin{table}[tbp] 
\centering
\caption[]{Radiative decay widths of baryons in keV. 
Systematic and statistical errors are added quadratically.} 
\label{key} 
\vspace{15pt}
\begin{tabular}{lccrcl}
\hline
& & & & & \\
$B \rightarrow B' + \gamma$ 
& \multicolumn{5}{c}{$\Gamma(B \rightarrow B' + \gamma)$} \\
& Ref.~\cite{lattice} & Ref.~\cite{Wagner} & Present & Exp. & \\ 
& & & & & \\
\hline
& & & & & \\
$\Sigma^{0} \rightarrow \Lambda + \gamma$ & & 
& 8.6 & $8.6 \pm 1.0$ & \protect\cite{Peterson} \\ 
$\Delta^{+} \rightarrow p + \gamma $ & $430 \pm 150$ & 350 
& 343.7 & $672 \pm 56$ & \protect\cite{PDG} \\ 
$\Delta^{0} \rightarrow n + \gamma $ & $430 \pm 150$ & 350 
& 341.5 & & \\ 
$\Sigma^{*,+} \rightarrow \Sigma^+ + \gamma$ & $100 \pm 26$ & 105 
& 140.7 & & \\
$\Sigma^{*,0} \rightarrow \Sigma^0 + \gamma$ & $17 \pm 4$ & 17.4 
&  33.9 & & \\
$\Sigma^{*,-} \rightarrow \Sigma^- + \gamma$ & $3.3 \pm 1.2$ & 3.6 
&   0.0 & & \\
$\Sigma^{*,0} \rightarrow \Lambda  + \gamma$ & & 265 
& 221.3 & & \\
$\Xi^{*,0} \rightarrow \Xi^0 + \gamma$ & $129 \pm 29$ & 172 
& 188.2 & & \\
$\Xi^{*,-} \rightarrow \Xi^- + \gamma$ & $3.8 \pm 1.2$ & 6.2 
&   0.0 & & \\
& & & & & \\
$\Lambda^*(1405) \rightarrow \Lambda + \gamma$ & & & 116.9  
& $ 27 \pm  8$ & \protect\cite{PDG} \\ 
$\Lambda^*(1405) \rightarrow \Sigma^{*,0} + \gamma$ & & & 0.0 & & \\
$\Lambda^*(1405) \rightarrow \Sigma^{0} + \gamma$ & & & 155.7 
& $ 10 \pm  4$ & \protect\cite{PDG} \\  
& & & & $ 23 \pm  7$ & \protect\cite{PDG} \\ 
$\Lambda^*(1520) \rightarrow \Lambda + \gamma$ & & & 85.1 
& $134\pm 23$ & \protect\cite{Landsberg,Mast} \\
& & & & $33\pm 11$ & \protect\cite{Bertini} \\ 
$\Lambda^*(1520) \rightarrow \Sigma^{*,0} + \gamma$ & & & 0.0 & & \\
$\Lambda^*(1520) \rightarrow \Sigma^{0} + \gamma$ & & & 180.4 
& $47\pm 17$ & \protect\cite{Bertini} \\ 
& & & & & \\
\hline
\end{tabular}
\end{table}

\clearpage

\clearpage

\clearpage

\clearpage

\clearpage

\clearpage

\clearpage

\clearpage

\end{document}